\documentclass[iop]{emulateapj}     
\newcommand{\tablepage}{}           

\shorttitle{Did the ancient Egyptians record Algol?}
\shortauthors{Jetsu et al.}
\begin{document}
\title{Did the ancient egyptians record the period of the 
       eclipsing binary \\ \object{Algol} -- the Raging one?}
\author{L. Jetsu} 
\author{S. Porceddu} 
\author{J. Lyytinen} 
\author{P. Kajatkari} 
\author{J. Lehtinen}
\author{T.Markkanen}
\affil{Department of Physics,
P.O. Box 64, FI-00014 University of Helsinki, Finland}
\email{lauri.jetsu@helsinki.fi}
\author{J. Toivari-Viitala}
\affil{Department of World Cultures,
P.O. Box 59, FI-00014 University of Helsinki, Finland}
\begin{abstract}
The eclipses in binary stars give precise
information of orbital period changes.
Goodricke discovered the 2.867 days period in the eclipses of \object{Algol} 
in the year 1783.
The irregular orbital period changes of this longest known
eclipsing binary continue to puzzle astronomers.
The mass transfer between the two members of this binary 
should cause a long-term increase of the orbital period, 
but observations over two centuries
have not confirmed this effect. 
Here, we present evidence indicating that the period of 
\object{Algol} was 2.850 days
three millennia ago. 
For religious reasons, the ancient Egyptians have recorded
this period into the Cairo Calendar,
which describes the repetitive changes of the Raging one.
Cairo Calendar may be the oldest preserved historical document
of the discovery of a variable star.
\end{abstract}

\section{Introduction} \label{introduction}

\tablepage
\begin{table*}
\begin{center}
\caption{CC prognoses for one Egyptian year. \label{tableone}}
\renewcommand{\arraystretch}{0.80}
\begin{scriptsize}
\begin{tabular}{rcccccccccccc}
\tableline\tableline
Day  & Akhet & Akhet & Akhet & Akhet & 
      Peret & Peret & Peret & Peret &
      Shemu & Shemu & Shemu & Shemu \\
    & I   & II  & III & IV  & I   & II  & III &  IV &  I  &  II & III & IV  \\
\tableline
$D$&$M=1$&$M=2$&$M=3$&$M=4$&$M=5$&$M=6$&$M=7$&$M=8$&$M=9$&$M=10$&$M=11$&$M=12$\\
\tableline 
 1 & GGG & GGG & GGG & GGG & GGG & GGG & GGG & GGG & GGG & GGG & GGG & GGG \\ 
 2 & GGG & GGG & --- & GGG & GGG & GGG & GGG & GGG & --- & --- & GGG & GGG \\ 
 3 & GGS & GGG & GGG & SSS & GGG & --- & --- & SSS & GGG & GGG & SSS & SSS \\ 
 4 & GGS & SGS & --- & GGG & GGG & GGG & GSS & GGG & SSS & SSS & GGG & SSG \\ 
 5 & GGG & SSS & --- & GGG & GSS & GGG & GGG & SSS & --- & GGG & SSS & GGG \\ 
 6 & SSG & GGG & GGG & SSS & GGG & --- & GGG & SSS & GGG & --- & --- & SSS \\ 
 7 & GGG & SSS & GGG & SSS & SSS & GGG & SSS & GGG & GGG & SSS & SSS & --- \\ 
 8 & GGS & GGG & --- & GGG & GGG & GGG & GGG & GGG & --- & GGG & SSS & GGG \\ 
 9 & GGG & GGG & SSS & GGG & GGG & GGG & GGG & --- & GGG & GGG & GGG & GGG \\ 
10 & GGG & GGG & GGG & GGG & SSS & SSS & SSS & --- & --- & GGG & SSS & GGG \\ 
11 & SSS & GGG & GGG & GGG & SSS & GGG & GGG & SSS & --- & SSS & SSS & SSS \\ 
12 & SSS & SSS & --- & SSS & --- & GGG & GGG & SSS & --- & GGG & --- & GGG \\ 
13 & GSS & GGG & SSS & GGG & GGG & SSS & GGG & SSS & --- & GGG & --- & GGG \\ 
14 & --- & GGG & SSS & GGG & SSS & SGG & --- & --- & --- & GGG & SSS & GGG \\ 
15 & GSS & GSS & SSS & --- & GGG & --- & SSS & GGG & --- & SSS & GGG & SSS \\ 
16 & SSS & GGG & GGG & GGG & GGG & --- & SSS & GGG & GGG & GGG & SSS & GGG \\ 
17 & SSS & GGG & --- & --- & SSS & GGG & SSS & SSS & GGG & SSS & --- & GGG \\ 
18 & GGG & SSS & SSS & SSS & GGG & SSS & GGG & --- & GGG & SSS & SSS & SSG \\ 
19 & GGG & GGG & SSS & SSS & SSS & GSS & --- & GGG & GGG & SSS & SSS & GGG \\ 
20 & SSS & SSS & SSS & SSS & SSS & SSS & SSS & --- & SSS & SSS & SSS & --- \\ 
21 & GGG & SSG & GGG & SSG & GGG & --- & --- & --- & SSS & SSG & GGG & GGG \\ 
22 & SSS & --- & --- & GGG & GGG & GGG & SSS & SSS & GGG & SSS & SSS & GGG \\ 
23 & SSS & --- & SSS & GGS & GGG & GGG & GGG & --- & GGG & GGG & SSS & SSS \\ 
24 & GGG & SSS & GGG & --- & GGG & SSS & SSS & SSS & --- & GGG & GGG & GGG \\ 
25 & GGS & SSS & GGG & --- & GGG & GGG & --- & SSS & GGG & GGG & GSG & GGG \\ 
26 & SSS & SSS & GGG & GGG & SSS & --- & SSS & --- & GGG & SSS & GGG & GSG \\ 
27 & GGG & SSS & GGG & GGS & GGG & --- & SSS & SSS & --- & SSS & SSS & SSS \\ 
28 & GGG & GGG & GGG & SSS & GGG & GGG & GGG & GGG & --- & GGG & SSS & GGG \\ 
29 & SGG & GGG & GGG & SSS & GGG & SSS & GGG & GGG & GGG & GGG & GGG & GGG \\ 
30 & GGG & GGG & GGG & GGG & GGG & SSS & GGG & GGG & GGG & GGG & GGG & GGG \\ \tableline
\end{tabular}
\end{scriptsize}
\addtolength{\tabcolsep}{+0.03cm}
\end{center}
\end{table*}

In Algol type eclipsing binaries (hereafter EB).
one member has evolved away from the main sequence and
Roche-lobe overflow 
has led to mass transfer (hereafter MT) 
to the other member.
MT can increase or decrease the orbital period 
$P_{\mathrm{orb}}$ \citep{Kwe58}.
Many EBs show only positive or negative $P_{\mathrm{orb}}$ changes.
Alternating period changes (hereafter APC)
seemed to occur only in EBs,
where one member displayed magnetic activity \citep{Hal89}.
Activity may explain APC \citep{App92}, but this phenomenon 
is still poorly understood \citep{Zav02,Lan06,Lia10}. 

Mon\-ta\-nari discovered \object{Algol} in 1669. 
It was the second variable discovered, 
73 years after the discovery of \object{Mira} by Fabricius.
\cite{Goo83} determined $P_{\mathrm{orb}}\!=\!2.^{\mathrm{d}}867$ of \object{Algol}
with naked eyes.
He received the Copley Medal for this outstanding achievement.
The observed (O) eclipses
can not be calculated (C) with a constant $P_{\mathrm{orb}}$. 
These $O\!-\!C$ show APC
cycles of 1.9, 32 and 180 years. 
\object{Algol} is actually a triple system.
The eclipsing stars in the $2.^{\mathrm{d}}867$ close orbit 
are Algol~A (B8~V) and Algol~B (K2~IV).
Algol~C (F1~IV) in the wide orbit
causes the 1.9 year cycle.
Applegate's theory may explain the longer cycles, 
because Algol~B has a convective envelope.
MT from Algol~B to Algol~A 
should cause a long-term $P_{\mathrm{orb}}$ increase,
but APC may have masked this effect \citep{Bie73}. 
This problem was discussed when \citet{Kis98} compared \object{Algol}
to \object{U~Cep}, where the parabolic $O\!-\!C$ trend
has confirmed a $P_{\mathrm{orb}}$ increase caused by MT. 
Evidence for this effect in \object{Algol}
is lacking after 230 years of observations.
Thus, any  $P_{\mathrm{orb}}$ information predating 1783~A.D. 
would be valuable.

Ancient Egyptian Scribes (hereafter AES) wrote Calendars of Lucky and Unlucky 
Days that assigned good and bad prognoses for the days of the year.
These prognoses were based on mythological and astronomical events 
considered influential for everyday life.
The best preserved calendar is the Cairo Calendar (hereafter CC)
in papyrus Cairo~86637 dated to 1271--1163 B.C. 
\citep{Bak66,Gle82,Lex75}.
Many CC prognoses had an astronomical origin,
because AES acting as ``hour--watchers'' 
observed bright stars for religious reasons during every clear night
\citep[e.g.][]{Lei89,Lei94,Kra02,Kra12}.
The traditions of AES in creating and copying tables of 
various different versions of star clocks spanned thousands of years. 
We have no exact knowledge about the volume of this activity and admittedly 
the evidence is scarce, but nevertheless the star clocks required 
existing astronomical observation practices. The little that we know about 
the observation practices comes mostly from Late Period (664--332 B.C.) 
sources such as the inscription on the statue of astronomer Harkhebi 
and the sighting instrument of Hor, son of Hor-wedja \citep{Cla95}.
\citet{Har02} argued that CC was a stellar almanac,
where known bright stars, like \object{$\alpha$ Car}, can be identified.
\citet[][Paper~I]{Por08}
detected the period of the Moon in CC.
Indications of a less significant period, $2.^{\mathrm{d}}85$, 
close to $P_{\mathrm{orb}}$ of \object{Algol}, were detected, 
but this connection had to be considered only tentative.
Here, we concentrate on statistics, astrophysics and astronomy. 
We show that $n \approx 200$ good prognoses
would induce $P_{\mathrm{Moon}}$ and $P_{\mathrm{Algol}}$ in CC,
even if the remaining $n \approx 700$ good and bad prognoses
had aperiodic origins \citep[][e.g. diseases, floods, feasts, winds]{Lei94}.
The connections between \object{Algol} and AES
are discussed in detail in {
Porceddu et al. (2013, in preparation, hereafter Paper~III),
where we date CC to 1224~B.C. 
A shift of $\pm 300^{\mathrm{y}}$
would not alter the main results presented here.

\tablepage
\begin{table}
\begin{center}
\caption{SSTP=1, 2, 3, ..., 23 and 24 created from Table \ref{tableone}. \label{tabletwo}}
\renewcommand{\arraystretch}{0.90}
\begin{tabular}{rrlrrrr}
\tableline\tableline
SSTP        & $N_0$ & ${\mathrm{Div}}$ & ${\mathrm{X}}$ & ${\mathrm{Remove}}$ & $n$ & $\Delta T$ \\ 
\tableline
  1 &               62 & Equation (\ref{adivision}) & {\rm G} & none & 564  & 359.3 \\
  2 &               62 & Equation (\ref{adivision}) & {\rm G} & $D=1$ & 528  & 358.3 \\
  3 &              187 & Equation (\ref{adivision}) & {\rm G} & none & 564  & 359.4 \\
  4 &              187 & Equation (\ref{adivision}) & {\rm G} & $D=1$ & 528  & 358.4 \\
  5 &              307 & Equation (\ref{adivision}) & {\rm G} & none & 564  & 359.3 \\
  6 &              307 & Equation (\ref{adivision}) & {\rm G} & $D=1$ & 528  & 358.3 \\
\tableline
  7 &               62 & Equation (\ref{bdivision}) & {\rm G} & none & 564  & 359.6 \\
  8 &               62 & Equation (\ref{bdivision}) & {\rm G} & $D=1$ & 528  & 358.6 \\
  9 &              187 & Equation (\ref{bdivision}) & {\rm G} & none & 564  & 359.6 \\
 10 &              187 & Equation (\ref{bdivision}) & {\rm G} & $D=1$ & 528  & 358.6 \\
 11 &              307 & Equation (\ref{bdivision}) & {\rm G} & none & 564  & 359.6 \\
 12 &              307 & Equation (\ref{bdivision}) & {\rm G} & $D=1$ & 528  & 358.6 \\
\tableline
 13 &               62 & Equation (\ref{adivision}) & {\rm S} & none & 351  & 354.0 \\
 14 &               62 & Equation (\ref{adivision}) & {\rm S} & $D=20$ & 321  & 354.0 \\
 15 &              187 & Equation (\ref{adivision}) & {\rm S} & none & 351  & 354.0 \\
 16 &              187 & Equation (\ref{adivision}) & {\rm S} & $D=20$ & 321  & 354.0 \\
 17 &              307 & Equation (\ref{adivision}) & {\rm S} & none & 351  & 354.0 \\
 18 &              307 & Equation (\ref{adivision}) & {\rm S} & $D=20$ & 321  & 354.0 \\
\tableline
 19 &               62 & Equation (\ref{bdivision}) & {\rm S} & none & 351  & 354.0 \\
 20 &               62 & Equation (\ref{bdivision}) & {\rm S} & $D=20$ & 321  & 354.0 \\
 21 &              187 & Equation (\ref{bdivision}) & {\rm S} & none & 351  & 354.0 \\
 22 &              187 & Equation (\ref{bdivision}) & {\rm S} & $D=20$ & 321  & 354.0 \\
 23 &              307 & Equation (\ref{bdivision}) & {\rm S} & none & 351  & 354.0 \\
 24 &              307 & Equation (\ref{bdivision}) & {\rm S} & $D=20$ & 321  & 354.0 \\
\tableline
\end{tabular}
\tablecomments{$N_0$ in Equation (\ref{gregorian}),
day division ${\mathrm{Div}}$ 
(Equations (\ref{adivision}) or (\ref{bdivision})),
selected prognoses (${\mathrm{X}}$),
removed prognoses (${\mathrm{Remove}}$), 
sample size ($n$)
and 
time span  ($\Delta T=t_n-t_1$).  }
\renewcommand{\arraystretch}{1.00}
\end{center}
\end{table}

\tablepage
\begin{table}
\begin{center}
\caption{Time points $t_i$ for all prognoses of Table 1. \label{tablethree}}
\begin{scriptsize}
\addtolength{\tabcolsep}{-0.12cm}
\renewcommand{\arraystretch}{0.90}
\begin{tabular}{rrrcrrrrrr}
\tableline\tableline
 & & & &\multicolumn{3}{c}{${\mathrm{Div}}$: Equation (\ref{adivision})}  &
        \multicolumn{3}{c}{${\mathrm{Div}}$:  Equation (\ref{bdivision})} \\
$D$&$M$ &$N_{\mathrm{E}}$& ${\mathrm{X}}$ &
$N_0\!=\!62$&$N_0\!=\!187$&$N_0\!=\!307$&$N_0\!=\!62$&$N_0\!=\!187$&$N_0\!=\!307$\\
   &    &          &           &
[days]  &[days]    &[days]     &[days]   & [days]  &[days]   \\
\tableline
 1 &  1 &   1 & G &   0.080 &   0.095 &   0.076 &   0.120 &   0.142 &   0.114\\
 1 &  1 &   1 & G &   0.239 &   0.284 &   0.227 &   0.359 &   0.426 &   0.341\\
 1 &  1 &   1 & G &   0.399 &   0.473 &   0.379 &   0.739 &   0.784 &   0.727\\
 2 &  1 &   2 & G &   1.080 &   1.095 &   1.076 &   1.120 &   1.142 &   1.113\\
 2 &  1 &   2 & G &   1.240 &   1.284 &   1.227 &   1.360 &   1.425 &   1.340\\
\end{tabular}
\addtolength{\tabcolsep}{+0.12cm}
\renewcommand{\arraystretch}{1.00}
\end{scriptsize}
\tablecomments{Table \ref{tablethree} is published in its entirety in the 
electronic edition of the {\it Astrophysical Journal Series}.  A portion is 
shown here for guidance regarding its form and content. 
\notetoeditor{Table 3 published Online}}
\end{center}
\end{table}

\section{Data} \label{data}

The CC prognoses are given in Table \ref{tableone}.
The ancient Egyptian year had 365 days. 
It contained 12 months ($M$) of 30 days ($D$).
Every month had 3 weeks of 10 days.
The year was divided into the flood
(Akhet), the winter (Peret) 
and the harvest (Shemu) seasons.
CC gave three prognoses a day,
except for the 5 additional ``epagomenal'' days of the year.
We use the German notation G=``gut''=``good'' and
S=``schlecht''=``bad'' \citep{Lei94}.
The notation for unreadable prognoses is ``--''. 
The Egyptian day began from dawn.
Daytime and nighttime were divided into 12 hours.
For example, GGS for ``I Akhet 25'' means
that the first two parts of this day were good,
but the third part was bad.
The logic of this day division procedure
has not been explained anywhere in the known Egyptian texts. 
The prognosis is usually the same, GGG or SSS, for the whole day.
However, 23 days have a heterogeneous prognosis, like GSS. 
\citet{Lei94} used the descriptions of such days
to infer how AES divided the day into parts. 
The first part refers to the morning, 
the second refers to mid--day and the third
refers to the evening, but may also include the night.

We computed Gregorian days ($N_{\mathrm{G}}\!=\!1\!\equiv\!$ Jan 1st) from
\begin{eqnarray}
\label{gregorian}
N_{\mathrm{G}} = 
\left\{
{\begin{array}{ll}  
 N_{\mathrm{E}} + N_0 - 1,               & N_{\mathrm{E}} \leq 366-N_0  \\
 N_{\mathrm{E}} + N_0 - 366,             & N_{\mathrm{E}} >    366-N_0, 
\end{array}  }
\right. 
\end{eqnarray}
{\noindent 
where $N_{\mathrm{E}} \!=\! 30 (M\!-\!1) \!+\! D$,
and $N_0\!=\!62,$ 187 or 307.
\citet{Lei94} has suggested $N_0\!=\!187$.
The values $N_0=307$ and 62 were obtained
by adding 120 and 240 days to $N_0=187$.
These three $N_0$ values were tested, 
because we did not know, where the Gregorian year began in CC.
We used
$\delta_{\odot} (N_{\mathrm{G}})$ $\! \approx \! - 23.45^{\mathrm{o}}$
$\cos{[360^{\mathrm{o}}(N_{\mathrm{G}}\!+\!10)/(365.25)]}$.
This accuracy was sufficient (see Section \ref{remarks}: 
11th paragraph).
The daytime at Middle Egypt ($\phi \!= \!26\arcdeg 41\arcmin$) 
was $l_{\mathrm{D}} (N_{\mathrm{G}})\!=\!$ 
$(24 /180\arcdeg)$ 
$\{{\mathrm{acos}} [ - \tan (\phi) \tan (\delta_{\odot}(N_{\mathrm{G}}))]\}$ hours.
Assuming that AES divided $l_{\mathrm{D}}(N_{\mathrm{G}})$ into three intervals gave
\begin{eqnarray}
t_1(N_{\mathrm{E}}) & = & 
(N_{\mathrm{E}}-1)+ (1/6) [l_{\mathrm{D}}(N_{\mathrm{G}})/24]  \nonumber \\
t_2(N_{\mathrm{E}}) & = & 
(N_{\mathrm{E}}-1)+ (3/6) [l_{\mathrm{D}}(N_{\mathrm{G}})/24]  \label{adivision} \\
t_3(N_{\mathrm{E}}) & = & 
(N_{\mathrm{E}}-1)+ (5/6) [l_{\mathrm{D}}(N_{\mathrm{G}})/24]. \nonumber 
\end{eqnarray}
\noindent
In our other alternative,
the daytime was divided into two intervals
and the nighttime was the third interval:
\begin{eqnarray}
t_1(N_{\mathrm{E}}) & = & 
(N_{\mathrm{E}}-1)+ (1/4) [l_{\mathrm{D}}(N_{\mathrm{G}})/24]  \nonumber \\
t_2(N_{\mathrm{E}}) & = & 
(N_{\mathrm{E}}-1)+ (3/4) [l_{\mathrm{D}}(N_{\mathrm{G}})/24]  \label{bdivision}  \\
t_3(N_{\mathrm{E}}) & = & 
(N_{\mathrm{E}}-1)+ 1/2 + (1/2)[l_{\mathrm{D}}(N_{\mathrm{G}})/24]. \nonumber 
\end{eqnarray}

{\noindent
These divisions represented the extremes that can be used
in placing three epochs within 24 hours.
We created 24 different samples of series of time points $t_i$
(hereafter SSTP) from Table \ref{tableone}.
The $t_i$ of G and S prognoses were separated.
The $D\!=\!1$ and 20 prognoses were always 
GGG and SSS (Table \ref{tableone}). 
We removed the $t_i$ of these days from some samples.
Table \ref{tabletwo} summarizes our SSTP.
The $t_i$ values for all prognoses are given in Table \ref{tablethree},
which is published only online.
Columns 1--4 give $D$, $M$, $N_E$ and the prognoses ${\mathrm{X}}$.
Columns 5--10 give $t_i$ calculated for different combinations 
of Equations (\ref{gregorian} -- {\ref{bdivision}).}

\section{Analysis} \label{analysis}

We did not analyse the ``bivalent data'' 
$y_i\!=\!y(t_i)\!=\!X(t_i) \!= \! {\mathrm{G ~or ~S.}}$
We analysed $t_i$, which fulfilled
$X(t_i)\!=\!{\mathrm{G}}$ or S. 
These ``circular data'' could be analysed with numerous
nonparametric methods \citep[e.g.][]{Bat81}.

\subsection{Rayleigh test} 

We analysed these $t_i$ with a Rayleigh test
between
$P_{\mathrm{min}}\!=\!1.^{\mathrm{d}}5$ 
and 
$P_{\mathrm{max}}\!=\!90^{\mathrm{d}}$.
Our $P_{\mathrm{min}}$ exceeded the data
spacing (Equations (\ref{adivision}) and (\ref{bdivision}))
and our $P_{\mathrm{max}}$ was $\Delta T/4$.
We have applied nonparametric methods to astronomical
\citep{Jet96a,Jet97b,Jet99b,Jet00a,Lyy02,Leh11,Leh12}
and geophysical data \citep{Jet97a,Jet00b,Lyy09}.
The sample size ($n$) and density ($\Delta T/(nP)$) 
of CC were better than in any of these previous studies. 

The phases are $\phi_i\!=\!{\mathrm{FRAC}}[(t_i-t_0)f],$
where ${\mathrm{FRAC}}[x]$ removes the integer part of $x$,
$f\!=\!P^{-1}$ is the tested frequency
and $t_0$ is an arbitrary epoch.
Rayleigh {\it test statistic} is 
$z(f) \! = \! |{\mathbf{R}}|^2/n$,
where $\theta_i=2 \pi \phi_i$,  ${\bf r}_i=[\cos{\theta_i},\sin{\theta_i}]$ 
and ${\bf R}\!=\!\sum_{i=1}^n {\bf r}_i$.
Rayleigh test null hypothesis is \\
$H_0$: {\it ``Phases $\phi_i$ calculated with an
arbitrary tested $P$ have a random distribution between 0 and 1.''} \\
If $H_0$ is true, 
${\bf r}_i$ point to random directions $\theta_i$ and $|{\bf R}| \!\approx\! 0$.
Coinciding $\theta_i$ give $|{\bf R}|\!=\!n$.
The probability density function is 
$f(z)\!\!=\!\!{\mathrm{e}}^{-z}$, which gives
$P(z \leq z_0)\!\!=\!\!F(z_0)\!\!=\!1\!-\!{\mathrm{e}}^{-z_0}$. 
If the tested $f$ are between $f_{\mathrm{min}}$ and $f_{\mathrm{max}}$,
the number of independent statistical tests is
$m\!=\!{\mathrm{INT}}[(f_{\mathrm{max}}-f_{\mathrm{min}})/f_0]$,
where ${\mathrm{INT}}[x]$ removes the decimal part of $x$ 
and $f_0\!=\!1/\Delta T$ 
\citep{Jet96b,Jet00b}.
The probability that $z(f)$ exceeds the value $z_0$ is 
\begin{equation}
Q = Q(z_0) = P(z(f) > z_0)= 1 - (1-{\mathrm{e}}^{-z_0})^{m}.
\label{critical}
\end{equation}
\noindent
This $Q$ is  
the {\it standard critical level}. We rejected $H_0$ if
\begin{equation}
Q < \gamma=0.001,
\label{reject}
\end{equation}
\noindent 
where $\gamma$ is called the {\it preassigned significance level.} 
We used simulations to check, if the 
above standard $Q$ estimates were reliable for the CC data.

\tablepage
\begin{table}
\begin{center}
\caption{Number of different daily prognosis combinations in 
Table 1. \label{tablefour}}
\renewcommand{\arraystretch}{0.90}
\begin{tabular}{crrrrrr}
\tableline\tableline
Prognosis   & \multicolumn{3}{c}{SSTP=1, 3, ..., 35} & \multicolumn{3}{c}{SSTP=2, 4, ..., 36} \\
combination & Days& {\rm G} & {\rm S} & Days& {\rm G} & {\rm S} \\          
\tableline
{\rm G}{\rm G}{\rm G} & 177 & 531 &   0 & 165 & 495 &   0 \\
{\rm G}{\rm G}{\rm S} &   6 &  12 &   6 &   6 &  12 &   6 \\
{\rm G}{\rm S}{\rm G} &   2 &   4 &   2 &   2 &   4 &   2 \\  
{\rm G}{\rm S}{\rm S} &   6 &   6 &  12 &   6 &   6 &  12 \\
{\rm S}{\rm S}{\rm S} & 105 &   0 & 315 &  95 &   0 & 285 \\
{\rm S}{\rm S}{\rm G} &   6 &   6 &  12 &   6 &   6 &  12 \\
{\rm S}{\rm G}{\rm G} &   2 &   4 &   2 &   2 &   4 &   2 \\
{\rm S}{\rm G}{\rm S} &   1 &   1 &   2 &   1 &   1 &   2 \\
\tableline    
Total     & 305 & 564 & 351 & 283 & 528 & 321 \\
\tableline    
``---''   &  55 &   0 &   0 &  53 &   0 &   0 \\
\tableline    
Total     & 360 & 564 & 351 & 336 & 528 & 321 \\
\tableline
          & \multicolumn{3}{c}{$\longrightarrow$ Table \ref{tablefive}} & 
            \multicolumn{3}{c}{$\longrightarrow$ Table \ref{tablesix}} \\
\tableline
\end{tabular}
\renewcommand{\arraystretch}{1.00}
\end{center}
\end{table}

\tablepage
\begin{table}
\begin{center}
\caption{Simulation of aperiodic data similar 
to SSTP=1, 3, ..., 23. \label{tablefive}}
\renewcommand{\arraystretch}{0.90}
\begin{tabular}{lr}
\tableline\tableline
Stage A: Event                                                 & 
P(Event)     \\
\tableline
${\mathrm{X}}^{\star}(t_1)={\mathrm{G}}$                                            & 
191/305      \\
${\mathrm{X}}^{\star}(t_1)={\mathrm{S}}$                                            & 
114/305      \\
\tableline
Stage B: Event                                                 & 
P(Event)     \\
\tableline
${\mathrm{X}}^{\star}(t_1)={\mathrm{G}} \Rightarrow {\mathrm{X}}^{\star}(t_2)={\mathrm{G}}$              & 
183/191      \\
${\mathrm{X}}^{\star}(t_1)={\mathrm{G}} \Rightarrow {\mathrm{X}}^{\star}(t_2)={\mathrm{S}}$              & 
8/191        \\
${\mathrm{X}}^{\star}(t_1)={\mathrm{S}} \Rightarrow {\mathrm{X}}^{\star}(t_2)={\mathrm{S}}$              & 
111/114      \\
${\mathrm{X}}^{\star}(t_1)={\mathrm{S}} \Rightarrow {\mathrm{X}}^{\star}(t_2)={\mathrm{G}}$              &   
3/114        \\
\tableline
Stage C: Event                                                 &
P(Event)     \\
\tableline
${\mathrm{X}}^{\star}(t_1)={\mathrm{G}}  {\rm ~and~}  {\mathrm{X}}^{\star}(t_2)={\mathrm{G}} \Rightarrow {\mathrm{X}}^{\star}(t_3)={\mathrm{G}}$ &
177/183      \\
${\mathrm{X}}^{\star}(t_1)={\mathrm{G}}  {\rm ~and~}  {\mathrm{X}}^{\star}(t_2)={\mathrm{G}} \Rightarrow {\mathrm{X}}^{\star}(t_3)={\mathrm{S}}$ &
6/183        \\
${\mathrm{X}}^{\star}(t_1)={\mathrm{G}}  {\rm ~and~}  {\mathrm{X}}^{\star}(t_2)={\mathrm{S}} \Rightarrow {\mathrm{X}}^{\star}(t_3)={\mathrm{G}}$ &
2/8          \\
${\mathrm{X}}^{\star}(t_1)={\mathrm{G}}  {\rm ~and~}  {\mathrm{X}}^{\star}(t_2)={\mathrm{S}} \Rightarrow {\mathrm{X}}^{\star}(t_3)={\mathrm{S}}$ &
6/8          \\
${\mathrm{X}}^{\star}(t_1)={\mathrm{S}}  {\rm ~and~}  {\mathrm{X}}^{\star}(t_2)={\mathrm{S}} \Rightarrow {\mathrm{X}}^{\star}(t_3)={\mathrm{S}}$ &
105/111      \\
${\mathrm{X}}^{\star}(t_1)={\mathrm{S}}  {\rm ~and~}  {\mathrm{X}}^{\star}(t_2)={\mathrm{S}} \Rightarrow {\mathrm{X}}^{\star}(t_3)={\mathrm{G}}$ &
6/111        \\
${\mathrm{X}}^{\star}(t_1)={\mathrm{S}}  {\rm ~and~}  {\mathrm{X}}^{\star}(t_2)={\mathrm{G}} \Rightarrow {\mathrm{X}}^{\star}(t_3)={\mathrm{G}}$ &
2/3          \\
${\mathrm{X}}^{\star}(t_1)={\mathrm{S}}  {\rm ~and~}  {\mathrm{X}}^{\star}(t_2)={\mathrm{G}} \Rightarrow {\mathrm{X}}^{\star}(t_3)={\mathrm{S}}$ &
1/3          \\
\tableline
\end{tabular}
\renewcommand{\arraystretch}{1.00}
\end{center}
\end{table}

\subsection{Simulation  of data similar to SSTP=1, 3, ..., 23}
\label{uneven}

Table \ref{tablefour} summarizes the real data:
number of ``Days'' (columns 2 and 5)
having the same ``Prognosis combination'' (column 1),
and number of individual ``G'' (columns 3 and 6) or
``S'' ~prognoses (columns 4 and  7).
For example,
the event ${\mathrm{X}}$$(t_1)\!=\!{\mathrm{G}}$ occurred with the probability 
of P(Event)$=\!(177\!+\!6\!+\!2\!+\!6)/305\!=\!191/305$
in the {\it real} data of SSTP=1,3,...,23.
The complementary event, ${\mathrm{X}}$$(t_1)\!=\!{\mathrm{S}}$, 
had P(Event)$=\!
(105\!+\!6\!+\!2\!+\!1)/305 \!=\!114/305$.
We {\it simulated} aperiodic data, where the prognosis
combinations of {\it real} data occurred with the same probabilities.
Table \ref{tablefour} (Columns 2--4) gave  
the probabilities P(event) of Table \ref{tablefive}.
Notations like ${\mathrm{X}}^{\star}(t_1)\!=\!{\mathrm{G}}$ 
or ${\mathrm{X}}^{\star}(t_1)\!=\!{\mathrm{S}}$ refer to the events 
that the {\it simulated} prognosis for the first time point 
$t_1$ of an arbitrary day is either G or S.
Aperiodic {\it simulated} data similar to the {\it real} data 
in SSTP=1,3,...,23 were generated with the following procedure:

\begin{enumerate}

\item We chose the simulated SSTP=1,3,... or 23.
The $t_1, t_2$ and $t_3$ for every $N_{\mathrm{E}}$ were calculated
with the $N_0$ and ${\mathrm{Div}}$ of this SSTP  (Table \ref{tabletwo}).
The time points of 55 randomly selected days were removed.

\item The random prognoses 
for each day were assigned using the probabilities P(Event) given in 
Table \ref{tablefive}. \\
-- Stage~A: The random prognosis ${\mathrm{X}}^{\star}(t_1)$ 
was assigned with the given probabilities P(event). \\
-- Stage B: The result for ${\mathrm{X}}^{\star}(t_1)$ 
then determined the probabilities
P(event) used in assigning ${\mathrm{X}}^{\star}(t_2)$. \\
-- Stage C:  The results for ${\mathrm{X}}^{\star}(t_1)$ 
and ${\mathrm{X}}^{\star}(t_2)$
then determined P(event) used in assigning ${\mathrm{X}}^{\star}(t_3)$.

\item We removed $t_i$ with 
${\mathrm{X}}^{\star}(t_i)=$ S for SSTP=1,3,...,11 and
$t_i$ with ${\mathrm{X}}^{\star}(t_i)=$ G for SSTP=13,15,...,23.

\end{enumerate}

\noindent
We used this procedure to simulate 10\,000 samples of
aperiodic random $t_i$ similar to every SSTP=1, 3, ..., 23.
This resembled the bootstrap approach \citep[e.g.][]{Jet96b},
because we created random samples imitating all the defects of the real data. 
Our repeated random sampling could also be called the Monte Carlo approach. 

The highest $z(f)$ peaks for the {\it real} data 
of SSTP=1 were at $P_1\!=\!29.^{\mathrm{d}}4$ and $P_2\!=\!2.^{\mathrm{d}}850$
(Figure \ref{figureone}a).
They reached 
$Q\!=\!0.0000034$ and 0.0012. 
Hence, $H_0$ should be rejected with
$P_1$, but not with $P_2$ (Equation (\ref{reject})). 

\begin{figure} 
\resizebox{\hsize}{!}{\includegraphics{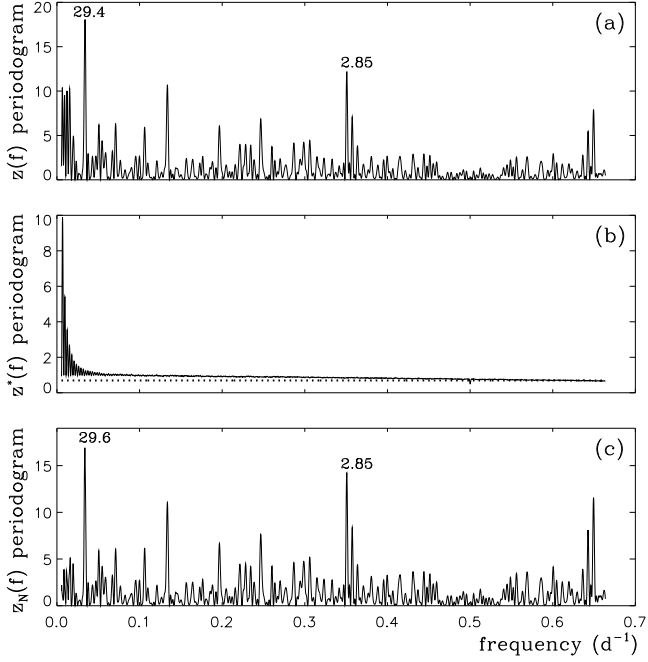}}
\caption{Three periodograms of SSTP=1. 
           (a) $f(z)$ for {\it real} data
                     gave $P_1\!=\!29.^{\mathrm{d}}4$ 
                     and $P_2\!=\!2.^{\mathrm{d}}85$.
           (b) $z^{\star}(f)$ (continuous line) for
                     {\it simulated} data similar to real data 
                     and level of 
               $z_0 \! = \!0.693 \! \equiv Q \! = \!0.5$ 
                     (dotted line).
           (c) $z_{\mathrm{N}}(f)$  for
                     {\it real} data gave $\!P_1\!=\!29.^{\mathrm{d}}6$ 
                                      and $\!P_2\!=\!2.^{\mathrm{d}}85$.
\label{figureone}}
\end{figure}

\begin{figure} 
\resizebox{\hsize}{!}{\includegraphics{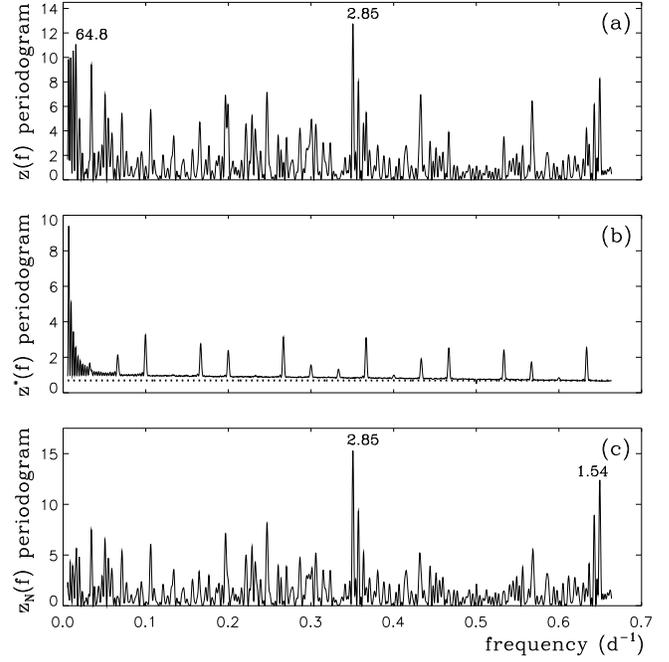}}
\caption{Three periodograms of SSTP=2. 
\label{figuretwo}}
\end{figure}

The {\it noise periodogram} $z^{\star}(f)$ for all 
10\,000 simulated aperiodic data samples similar to SSTP=1 
is shown in Figure \ref{figureone}b.
This $z^{\star}(f)$
is the {\it median}, not the {\it mean}, of $z(f)$ periodograms
for all 10\,000 simulated data samples at any particular $f$, 
because the probability density function of $z$ is not Gaussian.
This density function, ${\mathrm{e}}^{-z}$,  
predicts that half of the values fulfill $0 \le z \le 0.693$ 
and the rest fulfill $0.693 < z \le n$.
At the higher $f$, 
our $z^{\star}(f)$ approached $z_0\!=\!0.693 \!\equiv Q \!=\!0.5$
(Figure \ref{figureone}b: dotted line).
However, $z^{\star}(f)$ deviated 
from $z_0\!=\!0.693$ at lower $f$, 
i.e. the standard $Q$ estimates were not reliable. 
The $z^{\star}(f)$ periodogram displayed peaks at
\begin{equation} 
f=f(\Delta T,k)=[P(\Delta T, k)]^{-1}, 
\label{longperiods}
\end{equation}
where $k \geq 4$ was an integer 
and the long periods were $P(\Delta T,k)\!=\!{\Delta T/(k \!+\! 1/2)}$
(Figure \ref{figureone}b).
The $n\!=\!564$ simulated $t_i$ were 
nearly evenly spaced over $\Delta T\!=\!360^{\mathrm{d}}$ 
that contained $k\!+\!1/2$ cycles of $P(\Delta T,k)$.
The sum of ${\bf r}_i$ within the $k$ full cycles was close to zero.
The ${\bf r}_i$ within the remaining 1/2 cycle
pointed to same side of the unit circle.
This caused the $z^{\star}(f)$ peaks at $f(\Delta T,k)$.
These  $P(\Delta T,k)$ were unreal periods,
which gave us another reason for 
rejecting the use of the standard $Q$ estimates.

In the power spectrum analysis, 
the observed power at any tested $f$ is the 
``signal--power to noise--power ratio'' \citep{Sca82}.
We divided the standard Rayleigh test $z(f)$ periodogram 
for the real data with the noise periodogram $z^{\star}(f)$ 
for similar simulated aperiodic data.
This gave us the normalized periodogram
\begin{equation}
z_{\mathrm{N}}(f)= z(f)/ z^{\star}(f).
\label{normalization}
\end{equation}
To avoid any misunderstanding,
we emphasize that the power spectrum method was not applied here.
That parametric method relies on a sinusoidal model \citep{Sca82}.
It can be applied to a time series $y_i\!=\!y(t_i)$, 
but not to a series of time points $t_i$.
Rayleigh test is a nonparametric method.
There is no need to fit the data, because there is no model,
nor model parameters.

The highest $\!z_{\mathrm{N}}(f)$ peaks were at
$P_1\!\!=\!\!29.^{\mathrm{d}}6$ and $P_2\!\!=\!\!2.^{\mathrm{d}}850$ 
(Figure \ref{figureone}c).
Comparison to Figure \ref{figureone}a 
revealed that normalization shifted
$P_1$ from $29.^{\mathrm{d}}4$ to $29.^{\mathrm{d}}6$, 
but it did not shift $P_2$.
Normalization also eliminated numerous unreal peaks, 
especially in the lowest $f$ range.
We used $z_{\mathrm{N}}(f)$ to identify the best periods in SSTP=1.
Their significance had to be also solved from simulations,
because the standard $Q$ estimates were unreliable.
The peak at $f_1^{-1}=P_1=29.^{\mathrm{d}}6$
reached $z_1=z_{\mathrm{N}}(f_1)=16.68$.
We used Table \ref{tablefive}
to simulate ten million aperiodic
samples similar to the real data of SSTP=1
and calculated $z_{\mathrm{N}}(f)$ for each 
sample within $[f_1-f_0/2,f_1+f_0/2]$.
The highest peak satisfied $z_{\mathrm{N}}(f) > z_1=16.68$ 
only in 120 of these samples.
Hence, the {\it simulated critical level}
for $P_1=29.^{\mathrm{d}}6$  was $Q^{\star}=0.000012.$
For $P_2=2.^{\mathrm{d}}850$,
the same procedure gave $Q^{\star}=0.00014$, 
which fulfilled the criterion $Q^{\star}<\gamma=0.001$.
These changes differed, 
i.e.
$Q < Q^{\star}$ for $P_1=29.^{\mathrm{d}}6$ and 
$Q > Q^{\star}$ for $P_2=2.^{\mathrm{d}}850$.
We decided to revise the $H_0$ rejection criterion to
\begin{equation}
Q^{\star} < \gamma=0.001,
\label{newreject}
\end{equation}
\noindent
where $Q$ was substituted with $Q^{\star}$.
The periods $P_3\!=\!1.^{\mathrm{d}}5401$ $(Q^{\star}\!=\!0.00091)$ 
and $P_4\!=\!7.^{\mathrm{d}}48$ $(Q^{\star}\!=\!0.00091)$
also satisfied Equation (\ref{newreject}) for SSTP=1.
We will discuss these two unreal periods later.

\subsection{Simulation of data similar 
to SSTP=2,4,...,24  \label{even}} 

The sample size decreased after removing
the $D=1$ and 20 prognoses.
Columns 5--7 of Table \ref{tablefour} 
gave the P(Event) of Table \ref{tablesix}
used in generating simulated aperiodic data 
similar to the real data in SSTP=2,4,...,24.

\begin{enumerate}

\item We chose the simulated SSTP=2,4,... or 24.
The $t_1, t_2$ and $t_3$ for every $N_{\mathrm{E}}$ were calculated
with the $N_0$ and ${\mathrm{Div}}$ of this SSTP  (Table \ref{tabletwo}).
All time points at $D=1$ and 20 were first removed.
Then the time points of 53 randomly selected days were removed.

\item ${\mathrm{X}}^{\star}(t_1), {\mathrm{X}}^{\star}(t_2)$ 
and ${\mathrm{X}}^{\star}(t_3)$ were assigned as in Section \ref{uneven},
except that P(Event) were from Table \ref{tablesix}.

\item We removed $t_i$ with 
${\mathrm{X}}^{\star}(t_i)=$ S for SSTP=2,4,...,12 and
$t_i$ with ${\mathrm{X}}^{\star}(t_i)=$ G for SSTP=14,16,...,24.

\end{enumerate}

The highest $f(z)$ peaks for STTP=2
were at $P_1\!\!=\!\!2.^{\mathrm{d}}850$ and $P_2\!\!=\!64.^{\mathrm{d}}8$
(Figure \ref{figuretwo}a). 
Comparison between Figures \ref{figureone}b and \ref{figuretwo}b
revealed many new $z^{\star}(f)$ peaks caused by the
removal of $t_i$ at $D\!\!=\!\!1$.
The standard  $Q$ estimates were certainly unreliable. 
The highest $z_{\mathrm{N}}(f)$ peaks
were at 
$P_1\!=\!2.^{\mathrm{d}}850 ~(Q^{\star}\!=\!0.000094)$ 
and 
$P_2\!=\!1.^{\mathrm{d}}540 ~(Q^{\star}\!=\!0.00059)$ (Figure {\ref{figuretwo}}c).
Normalization did not shift $P_1$, 
but it revised $P_2$ from $64.^{\mathrm{d}}8$ to $1.^{\mathrm{d}}540$.
This unreal $P_2\!=\!P(\Delta T,k=5)\!=\!64.^{\mathrm{d}}8$
was predicted by Equation (\ref{longperiods}).
It was nicely eliminated by normalization.

The most striking {\it difference} between 
SSTP=1 and 2 was that the highly significant
$P_1=29.^{\mathrm{d}}6$ vanished.
The removal of ``GGG'' prognoses at $D=1$ caused this. 
The most important {\it similarity} was 
that the $2.^{\mathrm{d}}850$ period
fulfilled the criterion of Equation (\ref{newreject})
in both SSTP=1 and 2.
After removing the $D=1$ prognoses,
the significance of this  $2.^{\mathrm{d}}850$ period increased
($Q^{\star} \!= \!0.00014 \! \rightarrow \!0.000094$).
In conclusion, 
the removal of $D=1$ prognoses 
eliminated the $29.^{\mathrm{d}}6$ period and
the $2.^{\mathrm{d}}850$ period became the best period.
It also became more significant.

The $\sigma_{\mathrm{P}}$ estimates for all $P$ 
were determined from  $z_{\mathrm{N}}(f)$ 
with the bootstrap method \citep{Jet96b}.
Table \ref{tableseven} gives 
the best $P$ for G prognoses.
These $P$ satisfied the rejection criterion of Equation (\ref{newreject}).
All best $P$ for S prognoses (Table \ref{tableeight})
failed this criterion of Equation (\ref{newreject}).

\tablepage
\begin{table}
\begin{center}
\caption{Simulation of aperiodic data similar 
to SSTP=2,4,...,24. \label{tablesix}}
\renewcommand{\arraystretch}{0.90}
\begin{tabular}{lr}
\tableline\tableline
Stage A: Event                                         & 
P(Event)     \\
\tableline
${\mathrm{X}}^{\star}(t_1)={\mathrm{G}}$                                    & 
179/283      \\
${\mathrm{X}}^{\star}(t_1)={\mathrm{S}}$                                    & 
104/283      \\
\tableline
Stage B: Event                                         & 
P(Event)     \\
\tableline
${\mathrm{X}}^{\star}(t_1)={\mathrm{G}} \Rightarrow {\mathrm{X}}^{\star}(t_2)={\mathrm{G}}$      & 
171/179      \\
${\mathrm{X}}^{\star}(t_1)={\mathrm{G}} \Rightarrow {\mathrm{X}}^{\star}(t_2)={\mathrm{S}}$      & 
8/179        \\
${\mathrm{X}}^{\star}(t_1)={\mathrm{S}} \Rightarrow {\mathrm{X}}^{\star}(t_2)={\mathrm{S}}$      & 
101/104      \\
${\mathrm{X}}^{\star}(t_1)={\mathrm{S}} \Rightarrow {\mathrm{X}}^{\star}(t_2)={\mathrm{G}}$      &   
3/104        \\
\tableline
Stage C: Event                                                            &
P(Event)     \\  
\tableline
${\mathrm{X}}^{\star}(t_1)={\mathrm{G}}  {\rm ~and~}  {\mathrm{X}}^{\star}(t_2)={\mathrm{G}} \Rightarrow {\mathrm{X}}^{\star}(t_3)={\mathrm{G}}$ &
165/171      \\
${\mathrm{X}}^{\star}(t_1)={\mathrm{G}}  {\rm ~and~}  {\mathrm{X}}^{\star}(t_2)={\mathrm{G}} \Rightarrow {\mathrm{X}}^{\star}(t_3)={\mathrm{S}}$ &
6/171        \\
${\mathrm{X}}^{\star}(t_1)={\mathrm{G}}  {\rm ~and~}  {\mathrm{X}}^{\star}(t_2)={\mathrm{S}} \Rightarrow {\mathrm{X}}^{\star}(t_3)={\mathrm{G}}$ &
2/8          \\
${\mathrm{X}}^{\star}(t_1)={\mathrm{G}}  {\rm ~and~}  {\mathrm{X}}^{\star}(t_2)={\mathrm{S}} \Rightarrow {\mathrm{X}}^{\star}(t_3)={\mathrm{S}}$ &
6/8          \\
${\mathrm{X}}^{\star}(t_1)={\mathrm{S}}  {\rm ~and~}  {\mathrm{X}}^{\star}(t_2)={\mathrm{S}} \Rightarrow {\mathrm{X}}^{\star}(t_3)={\mathrm{S}}$ &
95/101      \\
${\mathrm{X}}^{\star}(t_1)={\mathrm{S}}  {\rm ~and~}  {\mathrm{X}}^{\star}(t_2)={\mathrm{S}} \Rightarrow {\mathrm{X}}^{\star}(t_3)={\mathrm{G}}$ &
6/101        \\
${\mathrm{X}}^{\star}(t_1)={\mathrm{S}}  {\rm ~and~}  {\mathrm{X}}^{\star}(t_2)={\mathrm{G}} \Rightarrow {\mathrm{X}}^{\star}(t_3)={\mathrm{G}}$ &
2/3          \\
${\mathrm{X}}^{\star}(t_1)={\mathrm{S}}  {\rm ~and~}  {\mathrm{X}}^{\star}(t_2)={\mathrm{G}} \Rightarrow {\mathrm{X}}^{\star}(t_3)={\mathrm{S}}$ &
1/3          \\
\tableline
\end{tabular}
\renewcommand{\arraystretch}{1.00}
\end{center}
\end{table}

\tablepage
\begin{table} 
\begin{center}
\caption{Best periods for the G prognoses. \label{tableseven}} 
\renewcommand{\arraystretch}{0.90}
\addtolength{\tabcolsep}{-0.10cm}
\begin{tabular}{cclcl}
\tableline\tableline
SSTP                                      &
$P$                       & $Q^{\star}$    &
$P$                       & $Q^{\star}$    \\
                                          &
[days]                    &               &
[days]                    &               \\
\tableline
1                                         &$
     29.6    \pm 0.2     $&$ 0.000012    $&$
      2.850  \pm 0.002   $&$ 0.00014     $\\
                                          &$
      1.5401 \pm 0.0008  $&$ 0.00091     $&$      
      7.48   \pm 0.02    $&$ 0.00091     $\\ \tableline  
2                                         &$
      2.850  \pm 0.002   $&$ 0.000094    $&$
      1.5400 \pm 0.0008  $&$ 0.00059     $\\ \tableline
3                                         &$
     29.6    \pm 0.2     $&$ 0.000015    $&$
      2.850  \pm 0.002   $&$ 0.00024     $\\
                                          &$
      1.5401 \pm 0.0008  $&$ 0.00057     $&$      
      \dots              $&$ \dots       $\\ \tableline
4                                         &$
      2.851  \pm 0.002   $&$ 0.00016     $&$
      1.5401 \pm 0.0008  $&$ 0.00037     $\\ \tableline
5                                         &$
      29.6    \pm 0.2     $&$ 0.000013   $&$
      2.851  \pm 0.002   $&$ 0.00015     $\\
                                          &$
      1.5404 \pm 0.0008  $&$ 0.00081     $&$      
      7.48   \pm 0.02    $&$ 0.00089     $\\ \tableline
6                                         &
$     2.851  \pm 0.002   $&$ 0.000094    $&$
      1.5401 \pm 0.0008  $&$ 0.00054     $\\
\tableline
7                                         &$
     29.5    \pm 0.2     $&$ 0.000012    $&$
      2.850  \pm 0.002   $&$ 0.00010     $\\
                                          &$
      7.48   \pm 0.02    $&$ 0.00079     $&$      
     \dots               $&$ \dots       $\\ \tableline
8                                         &
$     2.850  \pm 0.002   $&$ 0.000060    $&$
      \dots              $&$ \dots       $\\ \tableline
9                                         &$
     29.6    \pm 0.2     $&$ 0.000012    $&$
      2.851  \pm 0.002   $&$ 0.00016     $\\
                                          &$
      7.48   \pm 0.02    $&$ 0.00090     $&$      
     \dots               $&$ \dots       $\\ \tableline
10                                        &$
     2.851  \pm 0.002    $&$ 0.000096    $&$
      \dots              $&$ \dots       $\\ \tableline
11                                        &$
     29.6    \pm 0.2     $&$ 0.000013    $&$
      2.851  \pm 0.002   $&$ 0.000076    $\\
                                          &$
      7.48   \pm 0.02    $&$ 0.00088     $&$      
     \dots               $&$ \dots       $\\ \tableline
12                                        &
$     2.851  \pm 0.002   $&$ 0.000051    $&$
      \dots              $&$ \dots       $\\
\tableline
\end{tabular}
\addtolength{\tabcolsep}{+0.10cm}
\renewcommand{\arraystretch}{1.00}
\end{center}
\end{table}

\tablepage
\begin{table}
\begin{center}
\caption{Best periods for the S prognoses. \label{tableeight}}
\renewcommand{\arraystretch}{0.90}
\addtolength{\tabcolsep}{-0.00cm}
\begin{tabular}{cclcl}
\tableline\tableline
SSTP                                      &
$P$                      & $Q^{\star}$     &
$P$                      & $Q^{\star}$     \\
                                          &
[days]                    &               &
[days]                    &               \\
\tableline
13                                        &
$    30.3    \pm 0.2     $&$ 0.0021      $&$
      7.48   \pm 0.02    $&$ 0.0028      $\\ \tableline
14                                        &
$     2.795  \pm 0.003   $&$ 0.0066      $&$
      1.5570 \pm 0.0009  $&$ 0.0095      $\\  \tableline
15                                        &
$    30.3    \pm 0.2     $&$ 0.0018      $&$
      7.48   \pm 0.02    $&$ 0.0026      $\\  \tableline
16                                        &
$     2.822  \pm 0.002   $&$ 0.0079      $&$
      2.795  \pm 0.003   $&$ 0.0084      $\\  \tableline
17                                        &
$    30.3    \pm 0.2     $&$ 0.0020      $&$
      7.49   \pm 0.02    $&$ 0.0031      $\\  \tableline
18                                        &
$     2.796  \pm 0.002   $&$ 0.0053      $&$
      1.5487 \pm 0.0008  $&$ 0.011      $\\      
\tableline
19                                        &

$    30.2    \pm 0.2     $&$ 0.0018      $&$
      7.48   \pm 0.02    $&$ 0.0024      $\\  \tableline
20                                        &
$     2.795  \pm 0.003   $&$ 0.0053      $&$
      2.822  \pm 0.002   $&$ 0.011       $\\  \tableline
21                                        &
$    30.3    \pm 0.2     $&$ 0.0019      $&$
      7.48   \pm 0.02    $&$ 0.0028      $\\  \tableline
22                                        &
$     2.822  \pm 0.002   $&$ 0.0061      $&$      
      2.795  \pm 0.003   $&$ 0.0068      $\\  \tableline
23                                        &
$    30.3    \pm 0.2     $&$ 0.0021      $&$
      7.48   \pm 0.02    $&$ 0.0024      $\\  \tableline
24                                        &
$     2.795  \pm 0.003   $&$ 0.0041      $&$
      2.824  \pm 0.002   $&$ 0.012       $\\
\tableline
\end{tabular}
\addtolength{\tabcolsep}{+0.00cm}
\renewcommand{\arraystretch}{1.00}
\end{center}
\end{table}

\subsection{Results of the period analysis of all G prognoses
\label{allG}}

Four periods for G prognoses,
$29.^{\mathrm{d}}6$,
$2.^{\mathrm{d}}85$,
$1.^{\mathrm{d}}54$
and
$7.^{\mathrm{d}}48$,
satisfied rejection criterion of Equation (\ref{newreject}).

SSTP=3,5,7,9\&11 were similar to SSTP=1 in the sense 
that the G prognoses at $D\!=\!1$ were not removed. 
The two best periods for SSTP=3,5,7,9\&11
were within the error limits of the two best periods
$P_1=29.^{\mathrm{d}}6\pm0.^{\mathrm{d}}2$ 
and $P_2=2.^{\mathrm{d}}850 \pm 0.^{\mathrm{d}}002$ for SSTP=1.

SSTP=4,6,8,10\&12 were similar to SSTP=2, 
because the $D\!=\!1$ prognoses were removed. 
The best periods for these five SSTP were 
within the error limits of the best period 
$P_1=2.^{\mathrm{d}}850 \pm 0.^{\mathrm{d}}002$ for SSTP=2.

We compared the results for the SSTP=1\&2 pair in Section \ref{even}.
Comparison of the SSTP=3\&4, SSTP=5\&6, SSTP=7\&8, 
SSTP=9\&10 or SSTP=11\&12 pairs showed
that removing the G prognoses at $D=1$ always led to the same result:
{\it The best period $29.^{\mathrm{d}}6$  lost its significance, while
$2.^{\mathrm{d}}850$ became the new best period and the significance of
this periodicity increased.} 

The {\it unreal} $1.^{\mathrm{d}}54$ period detected in SSTP=1--6
was predicted by 
$P' (P_0, k_1, k_2)\! = \! [P^{-1}+(k_1/(k_2 P_0)]^{-1}$,
where 
$P=2.^{\mathrm{d}}85$ is the real period, 
$P_0=1.^{\mathrm{d}}0$ is the window period,
$k_1=-1$ and $k_2=1$ \citep{Tan48}.
This unreal period was not detected in SSTP=7--12,
because the daily $t_i$ were evenly distributed over 24 hours
with Equation (\ref{bdivision}) and therefore induced no
$P_0=1.^{\mathrm{d}}0$ window period.

We detected the {\it unreal} $7.^{\mathrm{d}}48$ period in 
SSTP=1,5,7,9\& 11,
but it vanished in SSTP=2,6,8,10\&12,
because it was nearly equal to $\delta t/4\!=\!7.^{\mathrm{d}}50$, 
where $\delta t\!=\!30^{\mathrm{d}}$ 
was the distance between the removed $D\!=\!1$ prognoses.

Normalization shifted the best period from
$29.^{\mathrm{d}}4$ to $29.^{\mathrm{d}}6$ in SSTP=1 
(Figures \ref{figureone}ac).
It caused similar shifts in SSTP=3,5,7,9\&11.
These shifts were always towards the mean of the synodic month, 
$P_{\mathrm{syn}}=29.^{\mathrm{d}}53$,
which would have been the most practical 
value for prediction purposes of AES.
The synodic month is not constant,
but varies between $29.^{\mathrm{d}}3$ and $29.^{\mathrm{d}}8$ 
in a year \citep{Ste91}.
Our $P_{\mathrm{Moon}}$ differed from $29.^{\mathrm{d}}53$ only by 
$+0.^{\mathrm d}07$ (SSTP=1,3,5,9\&11) and $-0.^{\mathrm d}03$ (SSTP=7).
We solved the precision
$\sigma_P$ that AES could have reached from $n$ observations of
$P_i\!\!=$ $\!\!P_{\mathrm{syn}} \!\!+ \!\!(A_{\mathrm{Moon}}/2) \sin{(2 \pi x_i)}$,
where $A_{\mathrm{Moon}}\!\!=\!\!0.^{\mathrm{d}}5$, 
$x_i\!\!=\!\!i /n$ and $i\!=\!1,...,n$.
It was
$\sigma_P^2\!\!= \!\![(A_{\mathrm{Moon}}/2)^2/n]$ $\! \sum_{i=1}^n [\sin{(iu)}]^2$, 
where $u \!=\! 2 \pi/n$. 
Then $\sum_{i=1}^n [\sin{(i u)}]^2\!=\!n/2\! - \!
[\cos{(n+1)} u \sin{(nu)}]/[2 \sin{u}]$ \citep{Gra94} 
gave $\sigma_P \!=\!2^{-3/2} A_{\mathrm{Moon}}\!=\!0.^{\mathrm{d}}18$.
This agreed with our $\sigma_{\mathrm{P}}\!=\!0.^{\mathrm{d}}2$ 
in Table \ref{tableseven}.
AES must have measured these changes for more than a year,
because their $P_{\mathrm{Moon}}$ estimate was much
closer to $29.^{\mathrm{d}}53$ than the expected 
observational $\pm 0.^{\mathrm{d}}2$ error. 

\subsection{Results of the period analysis of all S prognoses} 

There was no significant periodicity in S prognoses, because
all best $P$ failed the criterion of Equation (\ref{newreject}).
The $P\!=\!30.^{\mathrm{d}}3 \! \pm \! 0.^{\mathrm{d}}2$ 
and $7.^{\mathrm{d}}48 \! \pm \! 0.^{\mathrm{d}}02$
for SSTP=13,15,...,23
were the same within their error limits (Table \ref{tableeight}).
These $P$, originating from S at $D\!=\!20$,
were replaced by the new 
$P\!=\!1.^{\mathrm{d}}557, 2.^{\mathrm{d}}795$ and $2.^{\mathrm{d}}822$
for SSTP=14,16, ...,24.
The $2.^{\mathrm{d}}822$ and $2.^{\mathrm{d}}795$ periods were both 
close, but not equal to, the $2.^{\mathrm{d}}850$ period 
already detected from G prognoses.
$P'(P_0,k_1,k_2) \!\approx\! 1.^{\mathrm{d}}55$ 
was predicted by $P\!=\!2.^{\mathrm{d}}795$, 
$P_0\!=\!1.^{\mathrm{d}}0$, $k_1\!=\!1$ and $k_2\!=\!-1$ \citep{Tan48}.

\subsection{General remarks about the 
results  of period analysis} \label{remarks}

We analysed 24 different SSTP.
We had an infinite number of alternatives
for transforming Table \ref{tableone} into $t_i$,
but we simply could not invent any other realistic alternative 
transformations that would have altered our period analysis results.
For example, the available prognoses for $D\!=\!2$ were always ''GGG''. 
We performed additional tests, where $t_i$
at $D\!=\!1$ and 2 were removed. The best period was $2.^{\mathrm{d}}85$.
However, we could not test all possible alternatives
for removing $t_i$ from the data.

The unreal periods could be divided into two categories.
Those of the first category were present even in aperiodic data,
like the long periods predicted by Equation (\ref{longperiods}).
Normalization eliminated these first category unreal periods.
The second category unreal periods were induced by the real periods.
Some of these unreal periods could be predicted,
like the connection between the real $2.^{\mathrm{d}}85$ period
and the unreal $1.^{\mathrm{d}}54$ period \citep{Tan48}.
Normalization did not eliminate these second category unreal periods,
but they vanished when the real periodicity was removed,
like the unreal $7.^{\mathrm{d}}48$ period when the real $29.^{\mathrm{d}}6$
period was removed. All this indicated that only
the $29.^{\mathrm{d}}6$ and $2.^{\mathrm{d}}85$ periods were real. 

The standard $Q$ estimates were unreliable.
For example,
the evidence for $P_1\!\!=\!\!2.^{\mathrm{d}}850$ in SSTP=1 
was not indisputable, 
because it failed the criterion of Equation (\ref{reject}).
However, the significance of $P_1\!\!=\!\!2.^{\mathrm{d}}850$ 
was underestimated
$(Q\!\! > \!\!Q^{\star})$, while that 
of $P_2\!\!=\!\!29.^{\mathrm{d}}6$ was overestimated
$(Q\!\! < \!\!Q^{\star})$.
Periods $29.^{\mathrm{d}}6$ and $2.^{\mathrm{d}}850$
reached  $z_0\!\!=\!\!z_{\mathrm{N}}(f)\!\!=\!\!16.7$ and 14.6.
Inserting these $z_0$ into Equation (\ref{critical})
gave $Q(z_0)\!\!=\!\!0.000013$ and 0.00011.
This was nearly equal to $Q^{\star}\!\!=\!\!0.000012$ and 0.00014.
We emphasize that Equation (\ref{critical}) should 
not be applied to $z_{\mathrm{N}}(f)$.
However, using $z_0\!\!=\!\!z_{\mathrm{N}}(f)$ 
in Equation (\ref{critical}) gave $Q(z_0) \! \approx \! Q^{\star}$.
This indicated that our simulated statistics were robust.

Normalization allowed us to imitate the pattern of lucky and unlucky days, 
although we did not know the rules that were used to choose them.
It gave us the $Q^{\star}$ estimates 
and eliminated some of the unreal periods.
The best idea of all was to test what happens after
removing $P_{\mathrm{Moon}}$. 
This resulted in 
the $2.^{\mathrm{d}}850$ period being the only significant real period
and its significance increased.
CC does not give explicit clues as to why AES
assigned the prognoses with such regularity,
but the $2.^{\mathrm{d}}850$ period 
differs by $0.^{\mathrm{d}}017 \pm 0.^{\mathrm{d}}002$
from the current orbital period $2.^{\mathrm{d}}867328$ of \object{Algol}.
If this is indeed the reason for finding this periodicity in CC, 
then $P_{\mathrm{orb}}$ should have increased
about $25^{\mathrm{m}}$ since 1224~B.C.

Coinciding $\theta_i$ of $n_1$ periodic $t_i$ give
${\bf \!|A_1\!| } \!\!=\!$ $\!|\!\sum_{i=1}^{n_1} \!{\bf r}_i|\!\!=\!\!n_1$.
If aperiodic $t_i$ give
${\bf |A_2|}\!=\!| \sum_{i=1}^{n-n_1} {\bf r}_i | \! \approx \!0$,
then $z \!=$ $ \!{\bf |A_1} \!+\!{\bf A_2|}^2 / n \! \approx \! n_1^2/n$.
For example, $P\!=\!29.^{\mathrm{d}}4$ and $2.^{\mathrm{d}}851$
reached $z \! = \! 17.4$ and 12.1 in SSTP=1 ($n\!=\!564$),
which gave $n_1 \! \ge \! \sqrt{z n} \! \approx \! 99$ and 83. 
If 36 values 
connected to $P_{\mathrm{Moon}}$ were at $D\!=\!1$,
the other $t_i$ inducing this signal must have been
at $D\!=\!2$, 3, 29 or 30.
Thus, most of the G data could be aperiodic, 
because $n_1 \! \approx \! 200$ prognoses
could induce $P_{\mathrm{Moon}}$ and $P_{\mathrm{Algol}}$. 
The exact required number, $n_1$, can not be solved,
because the $\theta_i$ of all $n_1$ periodic $t_i$
can not be equal with 
Equations (\ref{adivision}) and (\ref{bdivision}).
There were 126 eclipses of \object{Algol} during $360^{\mathrm{d}}$.
If AES used {\it only one} G prognosis to 
mark each individual eclipse,
they may even have marked all eclipses into CC,
because reaching $n_1 \! \ga \!83$ requires interpolating many
of the $\approx 60$ daytime eclipses 
or of those eclipses that occurred when \object{Algol} 
was in conjunction with Sun. 

We used simulations to check, if a signal of $n_1$ periodic G
prognoses with $P_{\mathrm{Algol}}\!=\!2.^{\mathrm{d}}85$ 
would induce the $z(f)$ periodogram of Figure \ref{figuretwo}a.
We selected these $n_1$ periodic $t_i$ from the real data of SSTP=2. 
We assigned the remaining $n\!-\!n_1$ aperiodic random prognoses 
in such a way that the relative
number of different daily prognosis combinations
was the same as in the real data (Table \ref{tablefour}).
Our simulations reproduced the unreal $1.^{\mathrm{d}}54$ period,
as well as those predicted by Equation (\ref{longperiods}).
The period of
$P_{\mathrm{Algol}}\!\!=\!\!2.^{\mathrm{d}}85=\!\!57^{\mathrm{d}}/20$ induced
a $z(f)$ peak at $P_{\mathrm{Return}}\!\!=\!\!19^{\mathrm{d}}$ in many signals,
because a series of eclipses was repeated every 19 days
(see Figure \ref{figurefour}a: groups of vertical lines). 
AES may have noticed that eclipses ``returned'' 
exactly to the same epoch of the night
after $57^{\mathrm{d}}\!\!=\!\!3 \! \times \!19^{\mathrm{d}}$.
The relation
${{P_{\mathrm{Return}}} / {P_{k_3}}} \!- \! 
{{P_{\mathrm{Return}}} /  {P_{\mathrm{Algol}}}} \!=\! k_3 \!=\!\pm1, \pm 2, \dots$
predicted $z(f)$ peaks at $f\!=\!1/P_{k_3}$. 
For example, 
$P_{k_3=1}\!=\!3.^{\mathrm{d}}353$ and $P_{k_3=-1}\!=\!2.^{\mathrm{d}}478$
gave one cycle less or more than $P_{\mathrm{Algol}}$ during $P_{\mathrm{Return}}$. 
The $z(f)$ peaks at these unreal $P_{k_3}$ frequently 
exceeded the $P_{\mathrm{Algol}}$ peak
in weaker simulated signals ($n_1\!=\!40$).
However, $P_{\mathrm{Algol}}$ dominated over $P_{k_3}$ 
in stronger simulated signals ($n_1\!=\!100$).
When we divided the real STTP=1 and 2 data into two parts,
these unreal $P_{k_3}$ were weaker in the first part of CC
($N_{\mathrm{E}} \! \le \! 180$),
but many $P_{k_3}$ peaks 
exceeded that of $P_{\mathrm{Algol}}$ in the second part
($N_{\mathrm{E}} \!>\! 180$).
The real ($P_{\mathrm{Moon}}$ and $P_{\mathrm{Algol}}$)
and the unreal ($7.^{\mathrm{d}}48$ and $1.^{\mathrm{d}}54$)
periods were present in both parts.
Our simulations also revealed that 
the $z(f)\!=\!12.7$ peak at $f\!=\!1/P_{\mathrm{Algol}}$ 
in Figure \ref{figuretwo}a could be reproduced, 
if AES recorded only the observed, $n_1 \!\approx \!60$, 
nighttime eclipses by using
{\it more than one} G prognosis for each eclipse. 
AES may even have attributed importance to a connection 
between $P_{\mathrm{Moon}}$ and $P_{\mathrm{Algol}}$,
because $P_{\mathrm{Return}}\!=\!19^{\mathrm{d}}$
coincides with the difference
between $D\!=\!1$ (always GGG) and $D\!=\!20$ (always SSS) 
during {\it every} month.

The table from the Cosmology of Seti I and Ramses IV 
given in \citet[][pages 84--86]{Neu60} demonstrates how 
prone written documents from ancient Egypt were to writing errors. 
If we consider the amount of wrongly copied entries 
in the aforementioned table, 
it seems fair to test for an estimated 10\% of incorrect entries.
Therefore, we simulated periodic signals with $n_1\!=\!60$,
where six randomly chosen time points of each simulated signal were displaced.
These simulations revealed that if AES recorded only 
the observed $\approx 60$ yearly nighttime eclipses in CC,
the period of Algol could be discovered although 10\% 
of their entries were erroneous.

Here, we discuss our precision estimate, 
$\sigma_P\!=\!0.^{\mathrm{d}}002$, for $P_{\mathrm{Algol}}$.
The maximum separation between three 
$t_i$ {\it within} one day
is $8^{\mathrm{h}} \!\equiv \!\Delta \phi\!=\!0.12$ for $P_{\mathrm{Algol}}$.
An eclipse positioned to a correct third of the day,
had $\sigma_{\Delta \phi_i}\! \approx \!0.06$ for Equation (\ref{bdivision})
and less for Equation (\ref{adivision}). 
Our large samples contained four $t_i$ within each $P_{\mathrm{Algol}}$.
We obtained our $\sigma_P$ estimates for $P_{\mathrm{Moon}}$ and 
$P_{\mathrm{Algol}}$ from one year of data.
The long--term mean of the variable length of synodic month,
$P_{\mathrm{syn}}\!=\!29.^{\mathrm{d}}53$, 
was closer to our $P_{\mathrm{Moon}}$ than what
could be measured from only one year of observations
(Section \ref{allG}: last paragraph). 
This indicated that 
AES measured $P_{\mathrm{Moon}}$ changes over many years.
It was easier to measure 
long-term $P_{\mathrm{Algol}}$ than $P_{\mathrm{Moon}}$,
because the former remained practically constant.
A period change of $0.^{\mathrm{d}}017$ would revise 
the predictions radically,
because the current $P_{\mathrm{orb}}\!\!=\!\!2.^{\mathrm{d}}867$ 
predicts eclipses about $52^{\mathrm{h}}$ later in the end of a year
than $P_{\mathrm{orb}}\!=\!2.^{\mathrm{d}}850$.
The precision of $\sigma_P\!=\!0.^{\mathrm{d}}002$  
predicts correct nights for all yearly eclipses,
because the accumulated error is only $\pm 6^{\mathrm{h}}$. 

The ratio
$P_{\mathrm{Moon}}/P_{\mathrm{Algol}}
\!=\!29.^{\mathrm{d}}6/2.^{\mathrm{d}}850 \!=\!10.4$
was close to $P_{\mathrm{Week}}\!=\!10^{\mathrm{d}}$.
However, five facts contradicted the idea that
$P_{\mathrm{Week}}$ and $P_{\mathrm{Moon}}$ induced $P_{\mathrm{Algol}}$.
(1) There were no signs $P_{\mathrm{Week}}$ in CC.
(2) $P_{\rm Algol}\!=\!2.^{\mathrm{d}}850  \pm  0.^{\mathrm{d}}002$ 
was $55 \! \times \!\sigma_{P}$ smaller than $P_{\rm Moon}/10\!=\!2.^{\mathrm{d}}960$.
(3) After removing the $D=1$ prognoses, 
$P_{\rm Moon}$ and the unreal $7.^{\mathrm{d}}48$ period vanished,
but $P_{\rm Algol}$ {\it did not} vanish 
(compare Figures \ref{figureone}c and \ref{figuretwo}c).
Hence, $7.^{\mathrm{d}}48$ was connected to $P_{\rm Moon}$, 
but $P_{\rm Algol}$ was not. 
(4) After removing the $D=1$ prognoses, $P_{\mathrm{Moon}}$ vanished, 
but the significance of $P_{\rm Algol}$ {\it always} increased 
(Table \ref{tableseven}).
This indicated that $P_{\rm Algol}$ was not connected to $P_{\rm Moon}$. 
(5) The ratio $P_{\rm Moon}/P_{\rm Algol}$ 
induces a 0.4 phase difference in one month.
Events connected to $P_{\rm Moon}$ and $P_{\rm Algol}$ 
are totally out of phase throughout the whole year. 
Thus, $P_{\rm Moon}$ 
and/or $P_{\mathrm{week}}$ certainly did not induce $P_{\rm Algol}$.

It could be argued
that our test against $H_0$ was irrelevant,
because the data contained an algorithm.
The $z(f')$ peaks are at $f'$ that maximize $|{\bf R}|$.
The values of $|{\bf R}|$ or $f'$ do not depend on $H_0$,
but reveal any arbitrary $P'\!=\!1/f'$ coded with an algorithm,
e.g. $P_{\rm Moon}$ or $P_{\rm Algol}$.
The $z_0\!=\!z(f')$ value would give the $Q$ estimate
for $P'$ (Equation (\ref{critical})),
but we emphasized repeatedly that these standard $Q$ 
estimates were not valid.
We identified
the best periods $P'$  
from the $z_{\mathrm{N}}(f')$ peaks, 
which did not depend on $H_0$.
We solved the critical levels, $Q^{\star}$, from simulations,
which did not rely on $H_0$. 
In short, our period analysis results did not depend on $H_0$.

We also tested the constant daytime, 
$l_{\rm D}(N_{\mathrm{G}})\!=\!12^{\mathrm{h}}$, 
alternative for all $N_{\mathrm{G}}$ of the year
(Equations (\ref{adivision}) and (\ref{bdivision})).
The results did not change.
The $l_{\mathrm{D}}(N_{\mathrm{G}})$ changes,
as well as our $\delta_{\odot}(N_{\mathrm{G}})$ approximation in Section \ref{data},
had no influence on the results.


\section{Astrophysics} \label{astrophysics}

The physical parameters of \object{Algol} are given in Table \ref{tablenine},
where the subscripts ``1'' and ``2''
denote the ``A--B'' and ``AB--C'' systems.
The ZAMS masses 
were $m_{\mathrm{B}}=2.81 M_{\odot}$ and  $m_{\mathrm{A}}=2.50 M_{\odot}$ 
in the ``best--fitting'' evolutionary model of \citet{Sar93},
where Algol~B evolved away from the main sequence in 450 million years. 
This happened only a few million years ago.  
Roche--lobe overflow caused substantial MT to Algol~A,
which became more massive than Algol~B 
within less than 700\,000 years. 
MT is weaker at the current quiescent stage.
The complex APC of \object{Algol} may have ``masked'' \citep{Bie73} 
the presence of a small long--term $P_{\mathrm{orb}}$ 
increase that should have been 
observed as parabolic $O-C$ changes. 
MT from the less massive Algol~B to the more massive
Algol~A should lead to a long--term increase 
\begin{equation}
\dot{P}_{\mathrm{orb}}/P_{\mathrm{orb}} 
= - [3~ \dot{m}_{\mathrm{B}} ~(m_{\mathrm{A}}-m_{\mathrm{B}})]/(m_{\mathrm{A}} m_{\mathrm{B}})
\label{BH}
\end{equation}
\noindent 
where $\dot{P}_{\mathrm{orb}}$ is the rate of $P_{\mathrm{orb}}$ change,
$m_{\mathrm{A}}$ and $m_{\mathrm{B}}$ are the masses of the
gainer and the loser, and $\dot{m}_{\mathrm{B}}$ is the MT rate
\citep[][Eq. 5]{Kwe58}.
If $P_{\mathrm{orb}}$ was $2.^{\mathrm{d}}850$ in 1224~B.C.
and it has since then increased to $2.^{\mathrm{d}}867328$, 
constant $\dot{P}_{\mathrm{orb}}$ would give
$\dot{m}_{\mathrm{B}}\!=\! 
-2.2 \times 10^{-7} M_{\sun} {\rm ~per ~year}$.
This agreed with the ``best fitting'' evolutionary model
that predicted $\dot{m}_{\mathrm{B}}\!=\!-2.9 \times 10^{-7} M_{\odot}$ 
per year \citep{Sar93}. 
\citet{Sod80} noted that \object{Algol}'s MT
``is unlikely to be less than $10^{-7} M_{\odot} {\mathrm{~y}}^{-1}$''.
Constant MT is only an approximation,
because short MT bursts interrupt 
the long quiescent periods \citep[e.g.][]{Mal78}.
Equation (\ref{BH}) may also underestimate MT \citep{Zav02}.
However, more conservative MT estimates,
between $10^{-13}$ and  $10^{-8} M_{\odot} {\mathrm{~y}}^{-1}$,
have been published \citep{Har77,Cug77,Had84,Ric92}.

\cite{Bas00} discussed the accumulated long--term effects of 
Earth's non--uniform rotation to the O--C diagrams of EB.
Such effects also
shift the computed epochs of ancient solar eclipses \citep{Smi12}.
However, accumulated effects are insignificant within one year of data, like CC.
The days in 1224~B.C.
were $0.^{\mathrm{s}}055$ shorter than now,
because the increase has been about $0.^{\mathrm{s}}0017$ in 
a century \citep{Ste97}. 
If $P_{\mathrm{orb}}$ was $2.^{\mathrm{d}}850\,000$ 
in days in 1224~B.C.,
it would be $2.^{\mathrm{d}}849\,998$ in modern days.
This $0.^{\mathrm{d}}000\,002$ 
difference was 1\,000 times smaller than
our error $\sigma_P\!=\!0.^{\mathrm{d}}002$ for $2.^{\mathrm{d}}850$
and
8\,500 times smaller than the $0.^{\mathrm{d}}017$ period change.
Hence, Earth's non--uniform rotation did not prevent
a reliable comparison of the present-day $P_{\mathrm{orb}}$ of \object{Algol} 
to that in 1224~B.C. 

\tablepage
\begin{table}
\begin{center}
\caption{Physical parameters of the
Algol system \citep{Zav10}. \label{tablenine}}
\renewcommand{\arraystretch}{0.90}
\begin{tabular}{lll}
\tableline\tableline
Orbital elements              & 
Orbital elements              & 
Masses                        \\
A--B system                   & 
AB--C system                  & 
                              \\
\tableline
$a_1     =2.3 \pm 0.1 $        & 
$a_2= 93.8  \pm 0.2$           & 
$m_{\mathrm{A}}=3.7\pm0.2$       \\
$i_1     =98.6 $               & 
$i_2= 83.7\pm 0.1$             & 
$m_{\mathrm{B}}=0.8\pm0.1$       \\
$\Omega_1=7.4 \pm 5.2$         & 
$\Omega_2=132.7  \pm 0.1$      & 
$m_{\mathrm{C}}=1.5\pm0.1$       \\
$e_1     =0           $        & 
$e_2     = 0.225 \pm 0.005$    &  
                              \\
$P_1     =2.867328    $        & 
$P_2     =679.85 \pm 0.04$     &      
                               \\
\tableline
\end{tabular}
\renewcommand{\arraystretch}{1.00}
\tablecomments{
$[a_1]\!=\![a_2]\!=\arcsec/1000$, 
$[i_1]\!=\![i_2]\!=\![\Omega_1]\!=\![\Omega_2]\!=\!\arcdeg$, 
$[e_1]\!=\![e_2]\!=\!$ dimensionless,
$[P_1]\!=\![P_2]\!=\!{\rm d}$,
$[m_{\mathrm{A}}]\!=\![m_{\mathrm{B}}]\!=\![m_{\mathrm{C}}]=M_{\odot}$}
\end{center}
\end{table}

The perturbations of Algol~C are slowly changing $i_1$ 
and eclipses may not always occur.
\citet{Sod75} derived the period for these $i_1$ changes
\begin{eqnarray}
P_{i_1} \!=\!
{
{
 4 \left[1 \!+\!{{(m_{\mathrm{A}}+m_{\mathrm{B}}})/{m_{\mathrm{C}}}} \right]
{({P_2^2}/{P_1})}(1\!-\!e_2^2)^{3/2}
}
\over
{3
\left[{{(G_1}/{G_2)^2}}+{{2(G_1}/{G_2)}}\cos{\Psi}+1\right]^{1/2}\! \cos{\Psi}
}
},~~~
\label{soderhjelm}
\end{eqnarray}
\noindent
where
$\!G_1 \!\!= \!\! m_1 [G a_1 (1 \!-\!e_1)^2(m_{\mathrm{A}}\!+\!m_{\mathrm{B}})]^{1/2}$,
$m_1 \!\!= \!\! (m_{\mathrm{A}} m_{\mathrm{B}})/$ $(m_{\mathrm{A}}\!+\!m_{\mathrm{B}})$, 
$G_2 \!= \! m_2 [G a_2 (1 \!- \!e_2)^2(m_{\mathrm{A}}\!+\!m_{\mathrm{B}}\!+\!m_{\mathrm{C}})]^{1/2}$,
$m_2 \!= \! [(m_{\mathrm{A}} \!+\! m_{\mathrm{B}})m_{\mathrm{C}}]/
(m_{\mathrm{A}}\!+\!m_{\mathrm{B}}\!+\!m_{\mathrm{C}})$, 
$G$ is the gravitational constant and $\Psi$ is
the angle between the orbital planes of A--B and AB--C systems, which fulfills
\begin{eqnarray}
\cos{\Psi} = \cos{i_1} \cos{i_2} + \sin{i_1} \sin{i_2} \cos{(\Omega_1-\Omega_2)}.
\label{planeangle}
\end{eqnarray}
\noindent
Combining $\Psi \!=\! 95\arcdeg \!\pm\! 3\arcdeg$ \citep{Csi09} 
and $\Psi \!= \!86\arcdeg \!\pm\! 5\arcdeg$ \citep{Zav10}
to the values in Table \ref{tablenine} 
gave $P_{i_1}\!\!=\!\!25\,000$ and $31\,000$ years, i.e. $i_1$ may
have been stable during the past three millennia.
The $P_{i_1}$ lower limits were $14\,000$ and $16\,000$ years
for $\pm 1 \sigma_{\Phi}$.
Therefore, we could not confirm that eclipses occurred in 1224~B.C.

\tablepage
\begin{table}
\begin{center}
\caption{Thirteen variable star candidates not 
rejected with $C_1$ or $C_2$. \label{tableten}}
\addtolength{\tabcolsep}{-0.09cm}
\renewcommand{\arraystretch}{0.90}
\begin{tabular}{rcrrrrrrrr}
\tableline\tableline
Name                                       & $P$         & Type     &
$m_{\mathrm{max}}$                & $\Delta m$                          &
$\delta$                       & $a_{0}$    & $a_{30}$     & $a_{60}$  & $a_{\mathrm{max}}$  \\
                                            & [days]      &         &
[mag]                          & [mag]                              & 
[deg]                          & [h]        & [h]         &          [h] & [deg]    \\ 
\tableline
\object{$\zeta$ Pho}                        & 1.6697671          & EB &
3.91                           & 0.51                               & 
$-73$                          & 0     & 0  & 0   & $<\!0$    \\
\object{$\rho$ Per}                         & 50                 & SP  &
3.30                           & 0.70                               &
$+23$                          & 14    & 10  & 6 & 86   \\
\object{Algol}                               & 2.8673043          & EB  &
2.12                           & 1.27                               &
$+25$                          & 14    & 10 & 6    & 88   \\
\object{$\lambda$ Tau}                       & 3.9529478            & EB & 
3.37                           & 0.54                               &
$-1$                           & 12    & 8  & 4    & 62   \\
\object{$\mu$ Lep}                          & 2                  & CP  &
2.97                           & 0.44                               &
$-25$                          & 10    & 5  & 0   & 38   \\
\object{$\beta$ Dor}                        & 9.8426                & CE  &
3.46                           & 0.62                               &
$-65$                          & 0    &   0 & 0   &$<\!0$   \\ 
\object{$\zeta$)Gem}                        & 10.15073              & CE &
3.62                           & 0.56                               &
$+18$                          & 13    & 9  & 6    & 82 \\
\object{l Car}                              & 35.53584              & CE &
3.28                           & 0.90                               &
$-50$                          &  7    & 0  &  0   & 14   \\
\object{$\beta$ Lyr}                        & 12.913834          & EB  & 
3.25                           & 1.11                               &
$+34$                          & 15    & 10 & 6    & 82  \\
\object{R Lyr}                              & 46                 & SP   &
3.88                           & 1.12                               &
$+43$                          & 16    & 11 & 7    & 73  \\
\object{$\kappa$ Pav}                       & 9.09423            & CE &
3.91                           & 0.87                               &
$-60$                          & 4     & 0  &  0  & 3    \\
\object{$\eta$ Aql}                         & 7.176641               & CE  &
3.48                           & 0.91                               &
$-1$                           & 12   &  8  & 4  & 62    \\
\object{$\delta$ Cep}                       &  5.366341             & CE &
3.48                           & 0.89                               &
$+44$                          & 16   & 11  &  7 & 73 \\
\tableline
\end{tabular}
\addtolength{\tabcolsep}{0.09cm}
\renewcommand{\arraystretch}{1.00}
\tablecomments{Column 1: 
Name,
column 2: $P$,
column 3: Type 
(EB = eclipsing binary,
SP = semiregular pulsating star,  
CP = chemically peculiar
or
CE = cepheid),
columns 4 and 5: $m_{\mathrm{max}}$ and $\Delta m$,
column 6: $\delta$ in 1224~B.C,
columns 7--10: Time above altitudes 
$0\arcdeg$, $30\arcdeg$ and $60\arcdeg$
($a_{0},a_{30}$ and $a_{60}$)
and upper culmination ($a_{\mathrm{max}}$).}
\end{center}
\end{table}

\section{Astronomy} \label{astronomy}

Naked eye observers can discover periodicity 
in the Sun, the planets, the Moon and the stars.
Periods of the Sun and the planets exceed $90^{\mathrm{d}}$.
$P_{\mathrm{Moon}}$ was in CC.
Thus, the stars were the only other celestial objects,
where AES could have detected periodicity
between $1.^{\mathrm{d}}5$ and $90^{\mathrm{d}}$.
Here, we present eight criteria 
indicating that \object{Algol} was the most probable star, 
where AES could have discovered periodic variability.
General Catalogue of Variable Stars (hereafter GCVS\footnote{
GCVS at {\it http://www.sai.msu.su/groups/cluster/gcvs/gcvs/}
was accessed in 2008 November.}) gave
the maximum brightness ($m_{\mathrm{max}}$),
the amplitude ($\Delta m$) and the period ($P$) 
of all known over 40\,000 variables.
The criterion \\
$C_1$: {\it Variability fulfils  
$m_{\mathrm{max}} \leq 4.0$ and $\Delta m \geq 0.4$.} \\
gave those 109 stars, 
where variability could be discovered with 
naked eyes \citep[e.g.][]{Tur99}.
The criterion \\
$C_2$: {\it Period is known and fulfils $1.^{\mathrm{d}}5 \le P \le 90^{\mathrm{d}}$.} \\
left us with the 13 stars of Table \ref{tableten}. The criterion \\
$C_3$: {\it Variable was not below, or too close to, the horizon.} \\
eliminated \object{$\zeta$~Pho}, \object{$\beta$~Dor} 
and \object{$\kappa$~Pav}. The next criterion \\
$C_4$: {\it Variability can be predicted.} \\
eliminated \object{$\rho$ Per} \citep{Per93,Per96},
\object{$\mu$ Lep} \citep{Ren76,Hec87,Per87}
and \object{R Lyr} \citep{Per96,Per01}.
Their changes can not be predicted even today.
The changes of the remaining  
candidates are shown 
in Figures \ref{figurethree} and \ref{figurefour}. 
The light curves have not changed significantly since 
the discovery of these variables.
Therefore, we first modelled the light curves of the photometry in
\citet[][Algol]{Kim89},
\citet[][$\lambda$ Tau]{Gra59},
\citet[][$\zeta$ Gem, $\eta$ Aql and $\delta$ Cep]{Mof80},
\citet[][1 Car]{Dea77} and
\citet[][$\beta$ Lyr]{Asl87}.
We then obtained a full phase coverage by selecting 
a random sample of points from these models
and adding a Gaussian random $0.^{\mathrm{m}}01$ error to them.
This is what anyone would observe with an instrument having
a precision of $0.^{\mathrm{m}}01$. 
These curves could be descriptive,
because we were only interested in what can be detected with naked eyes.
\object{Algol} is easiest to observe with naked eyes. 
The next three criteria require that
the naked eye observer can {\it identify} suitable
and {\it eliminate} unsuitable comparison stars. 
The criterion \\
$C_5$: {\it Variability can be detected during a single night.} \\
eliminated \object{$\zeta$ Gem} and \object{l Car}.
The largest nightly changes of 
\object{$\beta$ Lyr},
\object{$\eta$ Aql}
and
\object{$\delta$ Cep}
were between $0.^{\mathrm{m}}2$ and $0.^{\mathrm{m}}4$.
The vertical lines in Figure \ref{figurefour}
show that the changes of 
these three could not be perceived during most nights.
The nightly changes of \object{Algol} and \object{$\lambda$ Tau},
$1.^{\mathrm{m}}27$ and $0.^{\mathrm{m}}54$, are the largest.
\object{Algol} is the only EB,
whose entire eclipse can be observed 
during a single night (Figure \ref{figurefour}h). 

\begin{figure} 
\resizebox{\hsize}{!}{\includegraphics{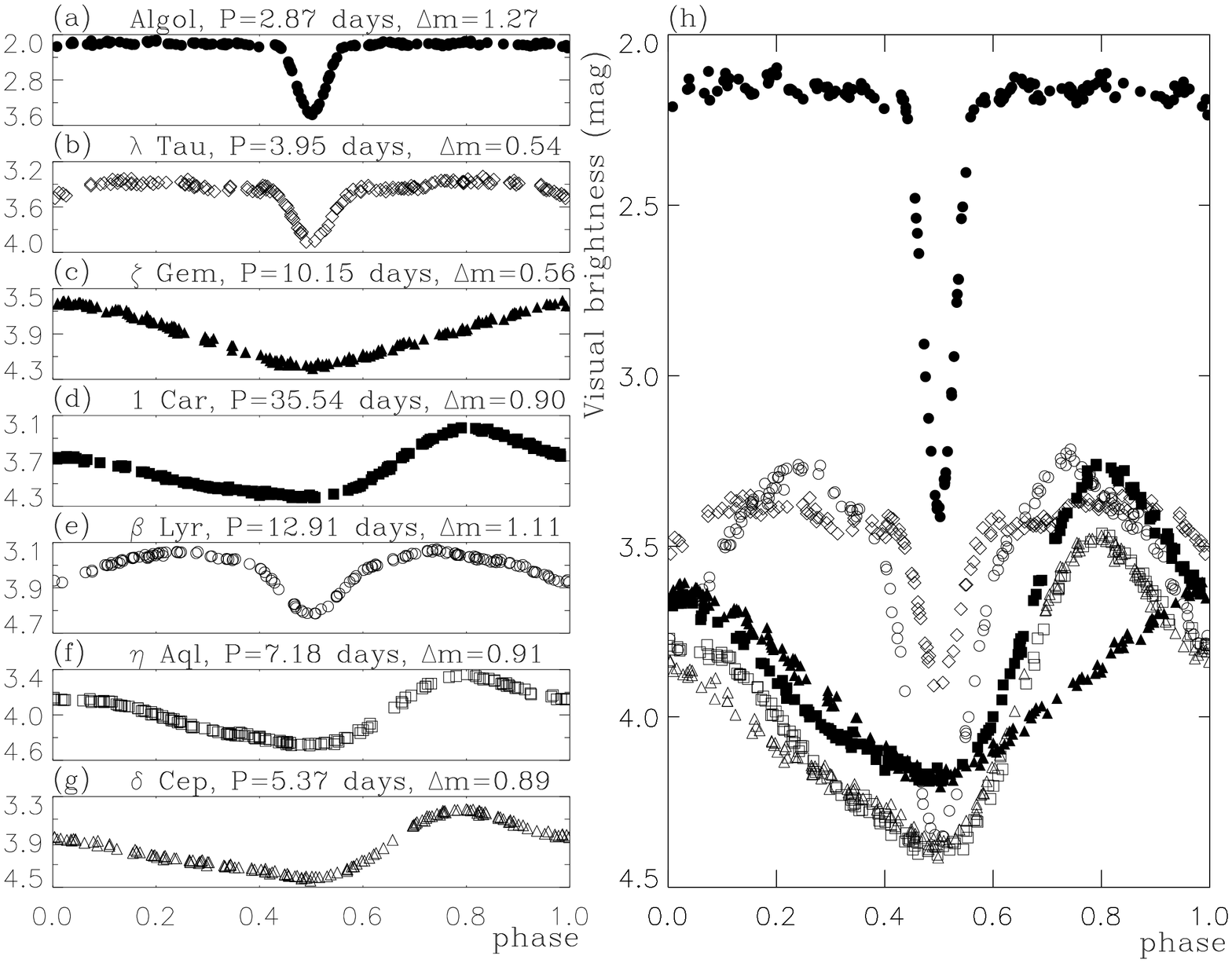}}
\caption{
Light curves of the best candidates as function of phase.
(a) Algol (closed circles),
(b) $\lambda$ Tau (open diamonds),
(c) $\zeta$ Gem (closed triangles),
(d) l Car (closed squares),
(e) $\beta$ Lyr (open circles),
(f) $\eta$ Aql (open squares)
and
(g) $\delta$ Cep (open triangles).
(h) All curves in the same scale: Algol is more than one magnitude
brighter than the other six variables and has the largest amplitude.
\label{figurethree}}
\end{figure}

\begin{figure} 
\resizebox{\hsize}{!}{\includegraphics{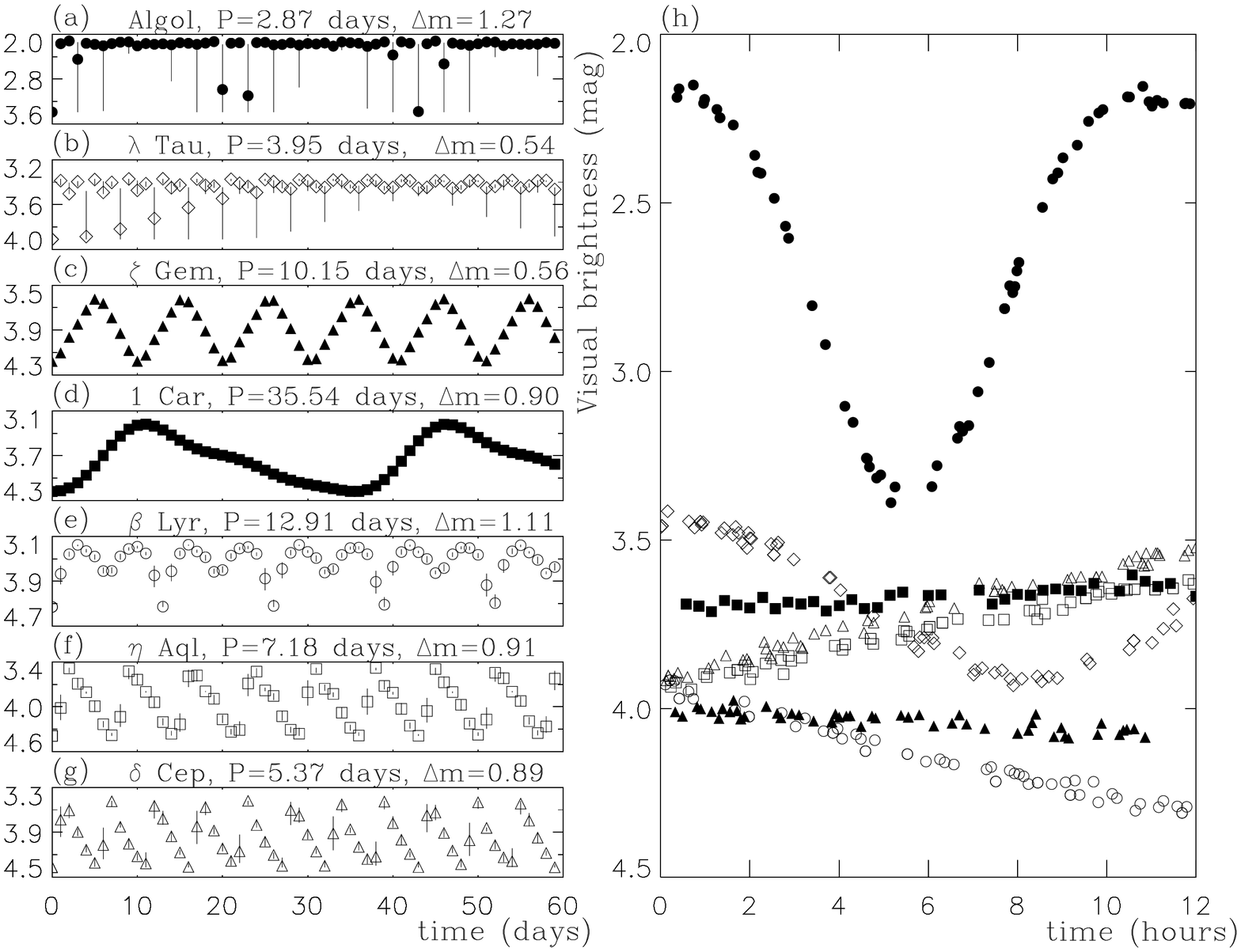}}
\caption{Light curves of the best candidates as a function of time.
(a--g) The symbols are the same as Figure \ref{figurethree}, 
but here they denote the brightness at mid--night.
The vertical lines (when visible) display the total range of the
brightness changes during a night lasting 12 hours.
(h) The nightly changes in the same scale.
The selected phase interval of each light curve is the one that would
induce the largest possible changes during a single night.
\label{figurefour}}
\end{figure}

\begin{figure*} 
\resizebox{\hsize}{!}{\includegraphics{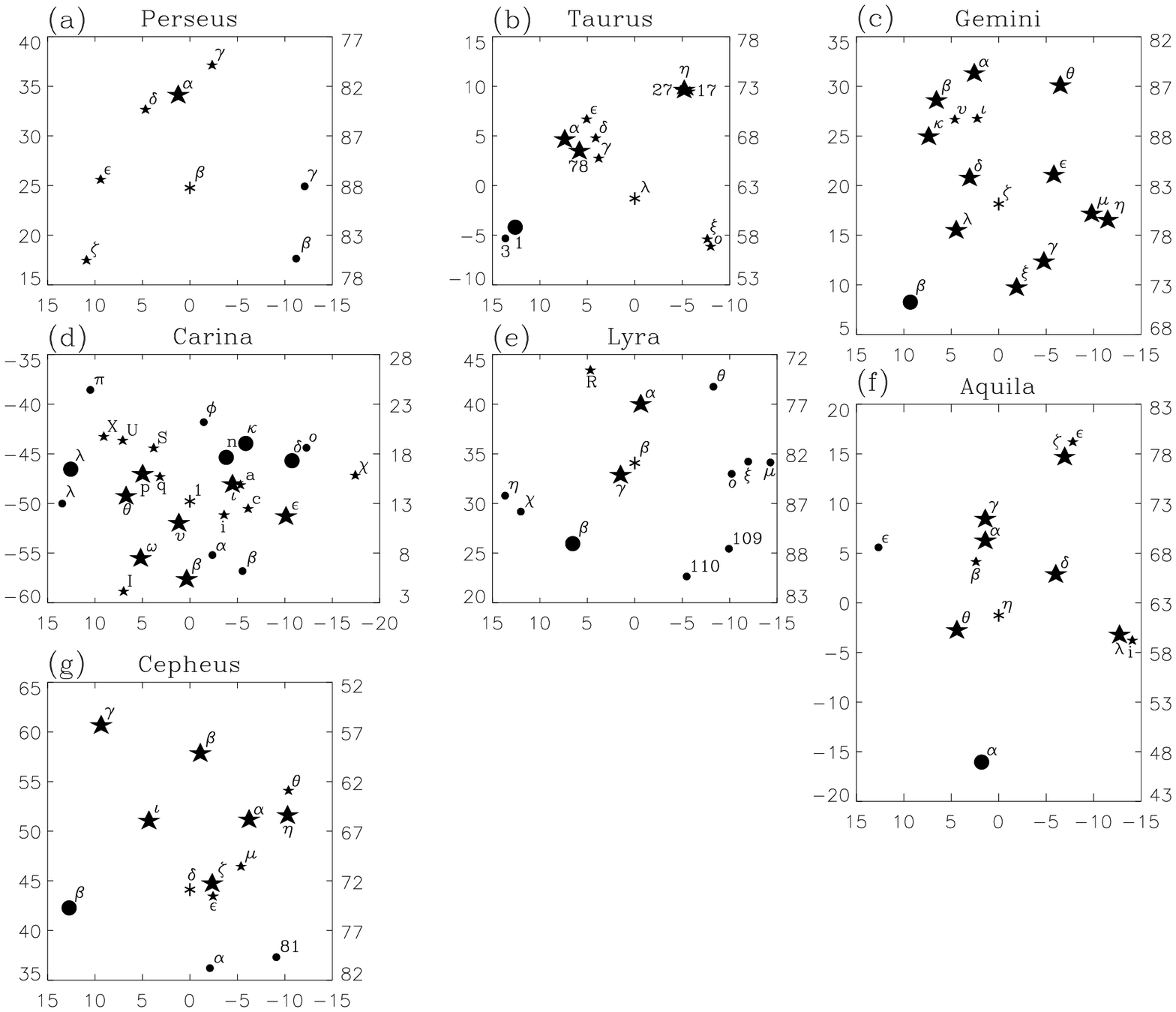}}
\caption{Stars close to the seven best candidates ({\normalsize $\!\!\ast$}).
The stars in the same modern constellation 
({\LARGE {$\star$}} and {\large{$\star$}})
are shown 
if $d \leq 20\arcdeg$ in Table \ref{tableeleven}. 
The stars not belonging to 
the same modern constellation  
({\LARGE $\bullet$} and {\normalsize{$\!\bullet$}})
are shown if $d \leq 15\arcdeg$. 
The right ascension (x-axis:$[\arcdeg]$), declination 
(left hand y-axis:$[\arcdeg]$) and
the altitude of upper culmination $a_{\mathrm{max}}$
(right hand y-axis:$[\arcdeg]$)
have been calculated for 1224~B.C.
\label{figurefive} \notetoeditor{Figure \ref{figurefive} published Online}}
\end{figure*}

We also checked how easy it is to discover
the changes of these seven variables in relation to nearby stars
({\normalsize $\!\!\ast$} in
online Figure \ref{figurefive} and online Table \ref{tableeleven}).
Our notations for brighter stars,
belonging and not belonging to
the same modern constellation as {\normalsize $\!\ast$},
are {\LARGE {$\star$}} and {\LARGE $\bullet$},
respectively \citep[][Bright Star Catalogue, hereafter BSC]{Hof91}.
The notations for comparison stars are
{\large{$\star$}} and {\normalsize{$\!\bullet$}}.
Table \ref{tableeleven} gives the Name, BSC number and
$m$ of stars close to $\ast$.
We give $\Delta m$, if it exceeds $0.^{\mathrm{m}}05$.
The distance from {\normalsize $\!\!\ast$} is 
$[d]\!=\!\arcdeg$. 
We compared four aspects:
(1) How many  {\LARGE {$\star$}} and {\LARGE $\bullet$}
outshined  {\normalsize $\!\!\ast$}?
(2) How many suitable and unsuitable 
{\large{$\star$}} and {\normalsize{$\!\bullet$}} were available?
(3) What kind of observations were required 
(Table \ref{tableten}: $P$, $\Delta m$)? 
(4) What were the extinction effects (Table \ref{tableten}: 
$a_0$, $a_{30}$, $a_{60}$)?
The criterion \\
$C_6$: {\it Variability changes the constellation pattern.} \\
eliminated all other candidates, 
except \object{Algol} and \object{$\lambda$ Tau}. 
It favoured \object{Algol}, the 2nd brightest star in a field,
where it fades below all six comparison stars in $5^{\mathrm{h}}$
(Figure \ref{figurefive}a).

\tablepage
\begin{table}
\begin{center}
\caption{Brightest stars close to the seven best variable star candidates.}
\addtolength{\tabcolsep}{0.0cm}
\begin{tabular}{cllccc}
\tableline\tableline
Notation &  Name           & Number  & $m$   & $\Delta m$ & $d~~$\\
         &                 &         & [mag] & [mag]      & [$\arcdeg$] \\
\tableline
{\LARGE {$\star$}} 
& $\alpha$    Per &  HR 1017 & 1.79 &  $-$       &  9.4 \\
{\normalsize $\ast$} 
& $\beta$     Per &  HR 936  & 2.12 & 1.27       &      \\
{\large $\star$} 
& $\zeta$     Per &  HR 1203 & 2.85 &  $-$       & 12.9 \\
{\large $\star$} 
& $\epsilon$  Per &  HR 1220 & 2.88 &  0.12      &  9.5 \\
{\large $\star$} 
& $\gamma$    Per &  HR 915  & 2.93 &  $-$       & 12.6 \\
\end{tabular}
\tablecomments{Table \ref{tableeleven} is published in its entirety in the 
electronic edition of the {\it Astrophysical Journal Series}.  A portion is 
shown here for guidance regarding its form and content. 
\notetoeditor{Table 11 published Online}}
\addtolength{\tabcolsep}{0.0cm}
\end{center}
\end{table}

Our previous criteria lead only to the discovery of {\it variability},
but not to the discovery of {\it periodicity}.
Discovery of periodicity in the smooth light curves of
\object{$\zeta$ Gem}, 
\object{l Car}, 
\object{$\eta$ Aql},
\object{$\delta$ Cep}
and 
\object{$\beta$ Lyr}
requires tabulation of differential magnitudes 
(i.e. a {\it time series}). 
Even if such tabulation had succeeded,
it is unlikely that AES could have used
a graphical solution to discover periodicity,
like Figures \ref{figurefour}c--g.
\object{Algol} and \object{$\lambda$ Tau} appear constant,
except during eclipses.
However, no {\it time series} is required to discover their periodicity,
but only a {\it series of time points.}
If the eclipse epochs are found to be 
multiples of the same number, then periodicity has been discovered.
The criterion \\
{\it $C_7$: Period of variability could be discovered by AES.} \\
did not eliminate \object{Algol} or \object{$\lambda$ Tau}.
The former is brighter with a larger $\Delta m$,
its exact eclipse epochs are easier to determine
and its altitude was higher in 1224~B.C.

The history of Astronomy
should indicate objectively the probability
for discovering variability and periodicity.
Fab\-ri\-cius (1596) discovered the first variable, \object{Mira}.
\object{Algol} was discovered next (Mon\-ta\-na\-ri, 1669).
\citet{Goo83} discovered its period.
\citet{Bax48} discovered the variability and 
$P_{\mathrm{orb}}$ of \object{$\lambda$ Tau},
but it took another 60 years to measure
the light curve due to the lack of suitable comparison stars \citep{Ste20}. 
The last criterion \\
{\it $C_8$: Variability and periodicity was discovered first. } \\
clearly favoured \object{Algol}.
Our eight criteria strongly indicated
that \object{Algol} was the most probable star,
where AES could have discovered periodic variability.

How could we constrain Algol's evolution,
if AES, Goodricke and modern astronomers used different magnitude systems? 
The time when the light fades, $t_i$, is the same in any system.
Hence, any $P$ inferred from these $t_i$ does not depend on the system.

\section{Discussion} \label{discussion}

AES were socially valued professionals, e.g in Astro\-nomy, Mathematics
and Medicine.
Their duties included also the measurement of time by observing stars
while they conducted the proper nightly rituals that
kept the Sun safe during its journey across the underworld
\citep{Lei89,Lei94,Har02,Kra02,Kra12}.
The timing of these rituals 
was important, because it had to appease the terrible guardians, 
who opened one gate of the underworld at each hour \citep{Cla89}.
The Sun was reborn at the 12th hour,
but only if AES performed the rituals absolutely right.
The risk that the Sun would never rise again was imminent.
With $P_{\mathrm{Algol}} =57^{\mathrm{d}}/20$, 
the eclipses always occur
exactly at the same modern hour after 57 nights.
Ancient Egyptian 
hours were of relative length so in winter the day hours were shorter
than in summer, and for night hours the reverse. Also due to the methods 
they employed to take into account the dawn and the dusk, 
their measurement of time was not precise \citep{Cla95}.
If an eclipse was observed in the end of the night,
the next eclipses occurred at three night intervals, 
but always about three and a half hours earlier, 
until they could not be observed at daytime. 
This sequence of nighttime eclipses was repeated every 19 days.
The eclipses also returned to the
same part of the night after 57 days.
These regularities occur with modern or ancient hours,
or to be precise, they could be discovered 
without any concept of hours.
Whatever \object{Algol} 
``did'' (blinked or not) on $D\! \approx \!1$ (always GGG), 
it always also did on $D\! \approx \! 20$ (always SSS).
There are nearly 300 clear nights a year in this area \citep{Mik95}.
Evidence of star clocks, which AES used to measure time from stars, 
spans over a millenium from the First Intermediate Period (ca. 2181-2055 B.C.) 
to the Late Period (664--332 B.C.). For this purpose they devised star tables 
to help with time keeping \citep{Cla95}. For example, the Ramesside star 
clocks contained thirteen rows of stars, where the first row stood for 
the opening of the night, the next eleven rows for the beginning of the 
consecutive hours of the night and the last row stood for the ending 
of the night.
AES must have encountered difficulties in the correct
identification of a very bright star (\object{Algol}), because is
was frequently outshined by six other dimmer nearby stars.
This star sometimes even lost and regained its brightness during
the same night (midnight eclipse).
Either they arrived at a known period value or
they just recorded the observed eclipses.
AES may have considered this 
variability to threaten the ``cosmic order''.
CC describes the repetitive transformation of the Eye of Horus, 
usually called ``Wedjat'' or 
``the Raging one'', from a peaceful to raging personality, 
with good or bad influence on the life of men \citep{Lei94}. 
A legend existed in which the enraged Eye of Horus nearly destroyed 
all mankind \citep{Lic76}. 
Most likely, AES linked \object{Algol}'s strange behaviour 
with this prominent legend.
It should be noted that in different contexts the concept of 
”the eye of Horus” could embody rather diverse meanings ranging 
from ritual equipment to even representing ``Re'', i.e. 
the Sun(god) \citep{Lei02b}.
It has also been argued that the Eye of Horus represents the Moon
\citep[e.g.][]{Set62,Lei94,Bel09}, and we discovered $P_{\mathrm{Moon}}$ in CC. 
However, the described repetitive changes of the Eye of Horus seem to
follow a much shorter time scale of a few days (Paper III: Section 2).

If AES recorded eclipses,
why are there no texts of \object{Algol} from other ancient cultures?
We argued that AES did not refer 
directly to \object{Algol} for religious reasons, 
but used indirect mythological references.
Half a year after our manuscript was submitted,
\citet{Smi12} showed that
AES also referred to solar eclipses only indirectly.
For example in the passage concerning III Peret 16, 
according to Leitz' calculations a New Moon day, one is forbidden 
to go outside and see the darkness \citep{Lei94}. The menacing presence 
of the god Seth over the morning of II Peret 14 has been believed to be a 
reference to the planet Mercury observed as a morning star \citep{Kra02}. 
Even the most direct astronomical descriptions from the ancient Egyptians 
such as the Cosmology of Seti I and Ramses IV do not plainly describe 
what happens in the sky but do that through mythological narrative 
\citep{Cla95}. 
This could explain the lack 
of references to the star itself.
There are indirect mythological
references to 
\object{Algol} also in other ancient cultures (Paper~III, Section 8).

The idea that CC contains significant new astrophysical information
may appear controversial.
A hypothesis is scientific only if it can be tested \citep[e.g.][]{Hem57}.
Scientific hypotheses are useful, if they give predictions based
on reasoning, like statistical tests or astrophysical relations. 
The word ``predict'' is used here
when extrapolating from the present--day to 1224~B.C. 
We use the present--day astrophysical parameters of \object{Algol} 
({\sc test i:} $P_{\mathrm{orb}}$, 
{\sc test ii}: $\dot{m}_{\mathrm{B}}$, 
{\sc test iii:} $\Phi$) and the present--day
astronomical catalogues ({\sc test iv:} GCVS, BSC). 

Two scientific hypotheses were tested.
We rejected our {\it statistical} hypothesis, $H_0$,
because the $29.^{\mathrm{d}}6$ and $2.^{\mathrm{d}}850$ periods were
indisputably detected with the new normalized Rayleigh test. 
This result was the core of our manuscript.

We applied four tests to our {\it astrophysical} hypothesis \\
$H_1$: {\it ``Period $2.^{\mathrm{d}}850$ in CC was $P_{\mathrm{orb}}$ of Algol.''}

{\sc test i:}  The present--day value is $P_{\mathrm{orb}}\!=\!2.^{\mathrm{d}}867$.
No one has presented evidence for $P_{\mathrm{orb}}$ increase
since \citet{Goo83} discovered this period.
An astrophysical relation (Equation (\ref{BH})) predicted
that MT from the less massive Algol~B to
the more massive Algol~A should have caused
such an increase \citep{Kwe58}. 
{\sc test i} supported $H_1$. 

{\sc test ii:} 
The present--day MT estimates
($[\dot{m}_{\mathrm{B}}]=M_{\odot}$ per year)
predicted the following $P_{\mathrm{orb}}$ values in 1224~B.C. \\ ~ \\
\citet{Har77}: $|\dot{m}_{\mathrm{B}}| \ge 10^{-9} 
\Rightarrow P_{\mathrm{orb}}\le2.^{\mathrm{d}}867$  \\
\citet{Cug77}: $|\dot{m}_{\mathrm{B}}| \approx 10^{-13} 
\Rightarrow P_{\mathrm{orb}}\le 2.^{\mathrm{d}}867$ \\
\citet{Sod80}: $|\dot{m}_{\mathrm{B}}| > 10^{-7} 
\Rightarrow P_{\mathrm{orb}}<2.^{\mathrm{d}}860$    \\
\citet{Had84}: $|\dot{m}_{\mathrm{B}}| \approx 10^{-8} 
\Rightarrow P_{\mathrm{orb}}\approx 2.^{\mathrm{d}}866$ \\
\citet{Ric92}: $10^{-11} \! \le \! |\dot{m}_{\mathrm{B}}| \! \le \! 10^{-10} 
\! \Rightarrow \! P_{\mathrm{orb}}\le2.^{\mathrm{d}}867$ \\
\citet{Sar93}: $|\dot{m}_{\mathrm{B}}| \approx 2.87 \times 10^{-7} 
\Rightarrow P_{\mathrm{orb}}\approx 2.^{\mathrm{d}}845$ \\ ~ \\
This large range, 
$\!10^{-13} \! \! \le \! \!|\dot{m}_{\mathrm{B}}| \!\! \le \!\! 2.87 \!\! \times \!\!10^{-7}$, 
gave no unique $\!P_{\mathrm{orb}}$ prediction.
However, these  $\!|\dot{m}_{\mathrm{B}}|$ 
were based on different approaches:
observations and models. 
The long quiescent periods
are sporadically interrupted by short bursts of MT.
All conservative MT estimates were based on observations  
\citep{Har77,Cug77,Ric92}, which may have coincided with
the long quiescent periods.
The bursts cause $P_{\mathrm{orb}}$ changes of several 
seconds in a year \citep[e.g.][]{Fri70,Mal78}.
MT in these bursts has to be much larger than our estimate,
$\!|\dot{m}_{\mathrm{B}}| \! = \!2.2 \times 10^{-7}$,
which predicts $P_{\mathrm{orb}}$ changes of only $0.^{\mathrm{s}}43$ in a year.
Our $\!|\dot{m}_{\mathrm{B}}|$ estimate, based on $\!H_1$,
may turn out to be valuable,
because many MT bursts must have occurred since 1224~B.C. 
{\sc test ii} did not contradict $H_1$. 

{\sc test iii:} 
A naked eye observer can determine $P_{\mathrm{orb}}$ 
from the present--day eclipses.
Eclipses have not necessarily occurred in all periods of history, 
because Algol~C changes $i_1$.
One argument against $H_1$ would have been that
the present--day $\Phi$ \citep{Csi09,Zav10} 
did not prove that eclipses occurred in 1224~B.C.
The astrophysical relations of Equations (\ref{soderhjelm})
and  (\ref{planeangle}) predicted this.
A few days {\it after} we submitted our manuscript,
\citet{Bar12} published a revised value,
$\Phi\!=\!90.2\arcdeg \pm 0.32\arcdeg$, which proved that 
eclipses similar to the present-day eclipses occurred also in 1224~B.C.
We could even argue that $H_1$ predicted their result. 
{\sc test iii} did not contradict $H_1$. 

{\sc test iv:} 
We searched for {\it all} celestial objects, 
where periodicity between $1.^{\mathrm{d}}5$ and $90^{\mathrm{d}}$
could be discovered with naked eyes.
$P_{\mathrm{Moon}}$ was in this range. 
The periods of the Sun and the planets were not.
We applied eight criteria to the
present--day data (GCVS, BSC) to eliminate all unsuitable variable stars.
The two most suitable remaining celestial 
objects were certainly the Moon and \object{Algol}.
We detected periodic signs of {\it only} these two celestial
objects in CC. {\sc test iv} supported $H_1$.  

{\sc tests} {\sc i\&iv} supported $H_1$.
{\sc tests} {\sc ii\&iii} did not contradict $H_1$, 
but indicated that $H_1$ could be true.
Thus, we could not prove that $H_1$ is definitely true.
Then again, no one from any field of science
has argued what {\it other} terrestrial or celestial phenomenon occurred 
regularly every third day, but {\it always} 3 hours and 36 minutes earlier 
than before, and caught the attention of AES? 

\section{Conclusions} \label{conclusions}

We discovered connections between \object{Algol} and 
AES writings that can hardly be a coincidence. 
All statistical, astrophysical, astronomical and egyptological details matched.
The period recorded in CC may represent a valuable constraint for future
studies of MT in EBs.
Goodricke's achievement in 1783 was outstanding.
The same achievement by AES, if true, 
was literally fabulous.

\acknowledgements
We thank 
M.A. Patricia Berg, 
Dr. Robert J. Demaree,
Ph.D. Heidi Jauhiainen,
prof. Karri Muinonen
and
prof. Heikki Oja 
for reviewing the original manuscript.
We thank Nadia Drake, Charles Choi, Anne Liljestr\"om 
and Stephen Battersby for encouragement received.
The SIMBAD database at CDS
and NASA's Astrophysics Data System (ADS) were used.
This work was supported by the Vilho, Yrj\"{o} and 
Kalle V\"{a}is\"{a}l\"{a} Foundation (P.K.),
the Finnish Graduate School in Astronomy and Space Physics (J.Le.)
and the Academy of Finland (J.T-V.). 

\bibliographystyle{apj}                                   
\bibliography{ApJjetsu}

\clearpage

\addtocounter{table}{-9}
\begin{table*}
\begin{center}
\caption{Time points $t_i$ for all prognoses of Table 1. }
\renewcommand{\arraystretch}{0.60}
\begin{tabular}{rrrcrrrrrr}
\tableline\tableline
 & & & &\multicolumn{3}{c}{${\mathrm{Div}}$: Equation (\ref{adivision})}  &
        \multicolumn{3}{c}{${\mathrm{Div}}$:  Equation (\ref{bdivision})} \\
$D$&$M$ &$N_{\mathrm{E}}$& ${\mathrm{X}}$ &
$N_0= 62$&$N_0=187$&$N_0=307$&$N_0= 62$&$N_0=187$&$N_0=307$\\
   &    &          &           &
(days)  &(days)    &(days)  &(days)   & (days)  &(days)   \\
\tableline
%
 1 &  1 &   1 & G &   0.080 &   0.095 &   0.076 &   0.120 &   0.142 &   0.114\\
 1 &  1 &   1 & G &   0.239 &   0.284 &   0.227 &   0.359 &   0.426 &   0.341\\
 1 &  1 &   1 & G &   0.399 &   0.473 &   0.379 &   0.739 &   0.784 &   0.727\\
 2 &  1 &   2 & G &   1.080 &   1.095 &   1.076 &   1.120 &   1.142 &   1.113\\
 2 &  1 &   2 & G &   1.240 &   1.284 &   1.227 &   1.360 &   1.425 &   1.340\\
\end{tabular}  
\end{center} 
\renewcommand{\arraystretch}{1.00}
\end{table*}          
\clearpage   
\begin{table*}
\renewcommand{\arraystretch}{0.60} 
\begin{center}
\begin{tabular}{rrrcrrrrrr}      \tableline
   &  & &  &\multicolumn{3}{c}{${\mathrm{Div}}$: Equation (\ref{adivision})}  &
            \multicolumn{3}{c}{${\mathrm{Div}}$:  Equation (\ref{bdivision})}  \\
$D$&$M$&$N_{\mathrm{E}}$&${\mathrm{X}}$&$N_0=62$&$N_0=187$&$N_0=307$            &
                             $N_0=62$&$N_0=187$&$N_0=307$           \\
   &    &        &          & (days) &(days)  &(days)               &
                              (days) &(days)  &(days)               \\  
\tableline
 1 &  2 &  31 & G &  30.085 &  30.091 &  30.072 &  30.128 &  30.137 &  30.108\\
 1 &  2 &  31 & G &  30.256 &  30.274 &  30.217 &  30.384 &  30.412 &  30.325\\
 1 &  2 &  31 & G &  30.427 &  30.457 &  30.362 &  30.756 &  30.774 &  30.717\\
 2 &  2 &  32 & G &  31.086 &  31.091 &  31.072 &  31.128 &  31.137 &  31.108\\
 2 &  2 &  32 & G &  31.257 &  31.274 &  31.217 &  31.385 &  31.411 &  31.325\\
\end{tabular}  
\end{center} 
\renewcommand{\arraystretch}{1.00}
\end{table*}          
\clearpage   
\begin{table*}
\renewcommand{\arraystretch}{0.60} 
\begin{center}
\begin{tabular}{rrrcrrrrrr}      \tableline
   &  & &  &\multicolumn{3}{c}{${\mathrm{Div}}$: Equation (\ref{adivision})}  &
            \multicolumn{3}{c}{${\mathrm{Div}}$:  Equation (\ref{bdivision})}  \\
$D$&$M$&$N_{\mathrm{E}}$&${\mathrm{X}}$&$N_0=62$&$N_0=187$&$N_0=307$            &
                             $N_0=62$&$N_0=187$&$N_0=307$           \\
   &    &        &          & (days) &(days)  &(days)               &
                              (days) &(days)  &(days)               \\  
\tableline
 1 &  3 &  61 & G &  60.091 &  60.086 &  60.072 &  60.136 &  60.130 &  60.108\\
 1 &  3 &  61 & G &  60.272 &  60.259 &  60.216 &  60.407 &  60.389 &  60.324\\
 1 &  3 &  61 & G &  60.453 &  60.432 &  60.360 &  60.772 &  60.759 &  60.716\\
 3 &  3 &  63 & G &  62.091 &  62.086 &  62.072 &  62.136 &  62.129 &  62.108\\
 3 &  3 &  63 & G &  62.273 &  62.258 &  62.216 &  62.409 &  62.388 &  62.324\\
\end{tabular}  
\end{center} 
\renewcommand{\arraystretch}{1.00}
\end{table*}          
\clearpage   
\begin{table*}
\renewcommand{\arraystretch}{0.60} 
\begin{center}
\begin{tabular}{rrrcrrrrrr}      \tableline
   &  & &  &\multicolumn{3}{c}{${\mathrm{Div}}$: Equation (\ref{adivision})}  &
            \multicolumn{3}{c}{${\mathrm{Div}}$:  Equation (\ref{bdivision})}  \\
$D$&$M$&$N_{\mathrm{E}}$&${\mathrm{X}}$&$N_0=62$&$N_0=187$&$N_0=307$            &
                             $N_0=62$&$N_0=187$&$N_0=307$           \\
   &    &        &          & (days) &(days)  &(days)               &
                              (days) &(days)  &(days)               \\  
\tableline
 1 &  4 &  91 & G &  90.094 &  90.081 &  90.075 &  90.141 &  90.121 &  90.112\\
 1 &  4 &  91 & G &  90.283 &  90.243 &  90.225 &  90.424 &  90.364 &  90.337\\
 1 &  4 &  91 & G &  90.471 &  90.404 &  90.374 &  90.783 &  90.743 &  90.725\\
 2 &  4 &  92 & G &  91.094 &  91.081 &  91.075 &  91.141 &  91.121 &  91.112\\
 2 &  4 &  92 & G &  91.283 &  91.242 &  91.225 &  91.424 &  91.363 &  91.337\\
\end{tabular}  
\end{center} 
\renewcommand{\arraystretch}{1.00}
\end{table*}          
\clearpage   
\begin{table*}
\renewcommand{\arraystretch}{0.60} 
\begin{center}
\begin{tabular}{rrrcrrrrrr}      \tableline
   &  & &  &\multicolumn{3}{c}{${\mathrm{Div}}$: Equation (\ref{adivision})}  &
            \multicolumn{3}{c}{${\mathrm{Div}}$:  Equation (\ref{bdivision})}  \\
$D$&$M$&$N_{\mathrm{E}}$&${\mathrm{X}}$&$N_0=62$&$N_0=187$&$N_0=307$            &
                             $N_0=62$&$N_0=187$&$N_0=307$           \\
   &    &        &          & (days) &(days)  &(days)               &
                              (days) &(days)  &(days)               \\  
\tableline
 1 &  5 & 121 & G & 120.095 & 120.076 & 120.080 & 120.142 & 120.114 & 120.120\\
 1 &  5 & 121 & G & 120.284 & 120.227 & 120.239 & 120.427 & 120.341 & 120.359\\
 1 &  5 & 121 & G & 120.474 & 120.379 & 120.399 & 120.784 & 120.727 & 120.739\\
 2 &  5 & 122 & G & 121.095 & 121.076 & 121.080 & 121.142 & 121.113 & 121.120\\
 2 &  5 & 122 & G & 121.284 & 121.227 & 121.240 & 121.427 & 121.340 & 121.360\\
\end{tabular}  
\end{center} 
\renewcommand{\arraystretch}{1.00}
\end{table*}          
\clearpage   
\begin{table*}
\renewcommand{\arraystretch}{0.60} 
\begin{center}
\begin{tabular}{rrrcrrrrrr}      \tableline
   &  & &  &\multicolumn{3}{c}{${\mathrm{Div}}$: Equation (\ref{adivision})}  &
            \multicolumn{3}{c}{${\mathrm{Div}}$:  Equation (\ref{bdivision})}  \\
$D$&$M$&$N_{\mathrm{E}}$&${\mathrm{X}}$&$N_0=62$&$N_0=187$&$N_0=307$            &
                             $N_0=62$&$N_0=187$&$N_0=307$           \\
   &    &        &          & (days) &(days)  &(days)               &
                              (days) &(days)  &(days)               \\  
\tableline
 1 &  6 & 151 & G & 150.092 & 150.072 & 150.085 & 150.138 & 150.108 & 150.128\\
 1 &  6 & 151 & G & 150.277 & 150.217 & 150.256 & 150.415 & 150.325 & 150.384\\
 1 &  6 & 151 & G & 150.461 & 150.362 & 150.427 & 150.777 & 150.717 & 150.756\\
 2 &  6 & 152 & G & 151.092 & 151.072 & 151.086 & 151.138 & 151.108 & 151.128\\
 2 &  6 & 152 & G & 151.276 & 151.217 & 151.257 & 151.414 & 151.325 & 151.385\\
\end{tabular}  
\end{center} 
\renewcommand{\arraystretch}{1.00}
\end{table*}          
\clearpage   
\begin{table*}
\renewcommand{\arraystretch}{0.60} 
\begin{center}
\begin{tabular}{rrrcrrrrrr}      \tableline
   &  & &  &\multicolumn{3}{c}{${\mathrm{Div}}$: Equation (\ref{adivision})}  &
            \multicolumn{3}{c}{${\mathrm{Div}}$:  Equation (\ref{bdivision})}  \\
$D$&$M$&$N_{\mathrm{E}}$&${\mathrm{X}}$&$N_0=62$&$N_0=187$&$N_0=307$            &
                             $N_0=62$&$N_0=187$&$N_0=307$           \\
   &    &        &          & (days) &(days)  &(days)               &
                              (days) &(days)  &(days)               \\  
\tableline
 1 &  7 & 181 & G & 180.087 & 180.072 & 180.091 & 180.131 & 180.108 & 180.136\\
 1 &  7 & 181 & G & 180.262 & 180.216 & 180.272 & 180.393 & 180.324 & 180.407\\
 1 &  7 & 181 & G & 180.437 & 180.360 & 180.453 & 180.762 & 180.716 & 180.772\\
 2 &  7 & 182 & G & 181.087 & 181.072 & 181.091 & 181.131 & 181.108 & 181.136\\
 2 &  7 & 182 & G & 181.262 & 181.216 & 181.272 & 181.392 & 181.324 & 181.408\\
\end{tabular}  
\end{center} 
\renewcommand{\arraystretch}{1.00}
\end{table*}          
\clearpage   
\begin{table*}
\renewcommand{\arraystretch}{0.60} 
\begin{center}
\begin{tabular}{rrrcrrrrrr}      \tableline
   &  & &  &\multicolumn{3}{c}{${\mathrm{Div}}$: Equation (\ref{adivision})}  &
            \multicolumn{3}{c}{${\mathrm{Div}}$:  Equation (\ref{bdivision})}  \\
$D$&$M$&$N_{\mathrm{E}}$&${\mathrm{X}}$&$N_0=62$&$N_0=187$&$N_0=307$            &
                             $N_0=62$&$N_0=187$&$N_0=307$           \\
   &    &        &          & (days) &(days)  &(days)               &
                              (days) &(days)  &(days)               \\  
\tableline
 1 &  8 & 211 & G & 210.082 & 210.075 & 210.094 & 210.123 & 210.112 & 210.141\\
 1 &  8 & 211 & G & 210.245 & 210.225 & 210.283 & 210.368 & 210.337 & 210.424\\
 1 &  8 & 211 & G & 210.409 & 210.374 & 210.471 & 210.745 & 210.725 & 210.783\\
 2 &  8 & 212 & G & 211.082 & 211.075 & 211.094 & 211.122 & 211.112 & 211.141\\
 2 &  8 & 212 & G & 211.245 & 211.225 & 211.283 & 211.367 & 211.337 & 211.424\\
\end{tabular}  
\end{center} 
\renewcommand{\arraystretch}{1.00}
\end{table*}          
\clearpage   
\begin{table*}
\renewcommand{\arraystretch}{0.60} 
\begin{center}
\begin{tabular}{rrrcrrrrrr}      \tableline
   &  & &  &\multicolumn{3}{c}{${\mathrm{Div}}$: Equation (\ref{adivision})}  &
            \multicolumn{3}{c}{${\mathrm{Div}}$:  Equation (\ref{bdivision})}  \\
$D$&$M$&$N_{\mathrm{E}}$&${\mathrm{X}}$&$N_0=62$&$N_0=187$&$N_0=307$            &
                             $N_0=62$&$N_0=187$&$N_0=307$           \\
   &    &        &          & (days) &(days)  &(days)               &
                              (days) &(days)  &(days)               \\  
\tableline
 1 &  9 & 241 & G & 240.077 & 240.080 & 240.095 & 240.115 & 240.120 & 240.142\\
 1 &  9 & 241 & G & 240.230 & 240.239 & 240.284 & 240.344 & 240.359 & 240.427\\
 1 &  9 & 241 & G & 240.383 & 240.399 & 240.474 & 240.730 & 240.739 & 240.784\\
 3 &  9 & 243 & G & 242.076 & 242.080 & 242.095 & 242.114 & 242.120 & 242.142\\
 3 &  9 & 243 & G & 242.229 & 242.240 & 242.284 & 242.343 & 242.361 & 242.426\\
\end{tabular}  
\end{center} 
\renewcommand{\arraystretch}{1.00}
\end{table*}          
\clearpage   
\begin{table*}
\renewcommand{\arraystretch}{0.60} 
\begin{center}
\begin{tabular}{rrrcrrrrrr}      \tableline
   &  & &  &\multicolumn{3}{c}{${\mathrm{Div}}$: Equation (\ref{adivision})}  &
            \multicolumn{3}{c}{${\mathrm{Div}}$:  Equation (\ref{bdivision})}  \\
$D$&$M$&$N_{\mathrm{E}}$&${\mathrm{X}}$&$N_0=62$&$N_0=187$&$N_0=307$            &
                             $N_0=62$&$N_0=187$&$N_0=307$           \\
   &    &        &          & (days) &(days)  &(days)               &
                              (days) &(days)  &(days)               \\  
\tableline
 1 & 10 & 271 & G & 270.073 & 270.085 & 270.092 & 270.109 & 270.128 & 270.138\\
 1 & 10 & 271 & G & 270.218 & 270.256 & 270.277 & 270.327 & 270.384 & 270.415\\
 1 & 10 & 271 & G & 270.364 & 270.427 & 270.461 & 270.718 & 270.756 & 270.777\\
 3 & 10 & 273 & G & 272.073 & 272.086 & 272.092 & 272.109 & 272.129 & 272.138\\
 3 & 10 & 273 & G & 272.218 & 272.257 & 272.276 & 272.326 & 272.386 & 272.414\\
\end{tabular}  
\end{center} 
\renewcommand{\arraystretch}{1.00}
\end{table*}          
\clearpage   
\begin{table*}
\renewcommand{\arraystretch}{0.60} 
\begin{center}
\begin{tabular}{rrrcrrrrrr}      \tableline
   &  & &  &\multicolumn{3}{c}{${\mathrm{Div}}$: Equation (\ref{adivision})}  &
            \multicolumn{3}{c}{${\mathrm{Div}}$:  Equation (\ref{bdivision})}  \\
$D$&$M$&$N_{\mathrm{E}}$&${\mathrm{X}}$&$N_0=62$&$N_0=187$&$N_0=307$            &
                             $N_0=62$&$N_0=187$&$N_0=307$           \\
   &    &        &          & (days) &(days)  &(days)               &
                              (days) &(days)  &(days)               \\  
\tableline
 1 & 11 & 301 & G & 300.072 & 300.091 & 300.087 & 300.108 & 300.136 & 300.131\\
 1 & 11 & 301 & G & 300.215 & 300.272 & 300.262 & 300.323 & 300.407 & 300.393\\
 1 & 11 & 301 & G & 300.359 & 300.453 & 300.437 & 300.715 & 300.772 & 300.762\\
 2 & 11 & 302 & G & 301.072 & 301.091 & 301.087 & 301.108 & 301.136 & 301.131\\
 2 & 11 & 302 & G & 301.215 & 301.272 & 301.262 & 301.323 & 301.408 & 301.392\\
\end{tabular}  
\end{center} 
\renewcommand{\arraystretch}{1.00}
\end{table*}          
\clearpage   
\begin{table*}
\renewcommand{\arraystretch}{0.60} 
\begin{center}
\begin{tabular}{rrrcrrrrrr}      \tableline
   &  & &  &\multicolumn{3}{c}{${\mathrm{Div}}$: Equation (\ref{adivision})}  &
            \multicolumn{3}{c}{${\mathrm{Div}}$:  Equation (\ref{bdivision})}  \\
$D$&$M$&$N_{\mathrm{E}}$&${\mathrm{X}}$&$N_0=62$&$N_0=187$&$N_0=307$            &
                             $N_0=62$&$N_0=187$&$N_0=307$           \\
   &    &        &          & (days) &(days)  &(days)               &
                              (days) &(days)  &(days)               \\  
\tableline
 1 & 12 & 331 & G & 330.074 & 330.094 & 330.082 & 330.111 & 330.141 & 330.123\\
 1 & 12 & 331 & G & 330.223 & 330.283 & 330.245 & 330.334 & 330.424 & 330.368\\
 1 & 12 & 331 & G & 330.371 & 330.471 & 330.409 & 330.723 & 330.783 & 330.745\\
 2 & 12 & 332 & G & 331.074 & 331.094 & 331.082 & 331.111 & 331.141 & 331.122\\
 2 & 12 & 332 & G & 331.223 & 331.283 & 331.245 & 331.334 & 331.424 & 331.367\\
\end{tabular}
\tablecomments{Columns 5--10 have been calculated with $N_0=62,$ 
187 or 307 in Equation (\ref{gregorian})
combined with the day divisions ${\mathrm{Div}}$: 
Equation (\ref{adivision}) or (\ref{bdivision}).}
\end{center}
\renewcommand{\arraystretch}{1.00}
\end{table*}

\clearpage

\tablepage
\addtocounter{table}{7} 
\begin{table}
\begin{center}
\caption{Brightest stars close to the seven best variable star candidates. 
\label{tableeleven}}
\renewcommand{\arraystretch}{1.00}
\addtolength{\tabcolsep}{0.0cm}
\begin{scriptsize}
\begin{tabular}{cclccc}
\tableline\tableline
Notation &  Name            &            & $m$   & $\Delta m$ & $d~~$\\
         &                  &            & [mag] & [mag]      & [$\arcdeg$] \\
\tableline
{\LARGE {$\star$}} 
& $\alpha$    Per &  HR 1017 & 1.79 &  $-$       &  9.4 \\
{\normalsize $\ast$} 
& $\beta$     Per &  HR 936  & 2.12 & 1.27       &      \\
{\large $\star$} 
& $\zeta$     Per &  HR 1203 & 2.85 &  $-$       & 12.9 \\
{\large $\star$} 
& $\epsilon$  Per &  HR 1220 & 2.88 &  0.12      &  9.5 \\
{\large $\star$} 
& $\gamma$    Per &  HR 915  & 2.93 &  $-$       & 12.6 \\
{\large $\star$} 
& $\delta$    Per &  HR 1122 & 3.01 &  $-$       &  9.2 \\
{\normalsize {$\bullet$}}
& $\gamma$    And &  HR 603  & 2.26 &  $-$       & 12.1 \\
{\normalsize {$\bullet$}}
& $\beta$     Tri &  HR 622  & 3.00 &  $-$       & 13.1 \\
\tableline
{\LARGE {$\star$}} 
& $\alpha$    Tau &  HR 1457 & 0.75 &  0.20      &  9.4 \\
{\LARGE {$\star$}} 
& $\beta$     Tau &  HR 1791 & 1.65 &  $-$       & 25.7 \\
{\LARGE {$\star$}} 
& $\eta$      Tau &  HR 1165 & 2.87 &  $-$       & 12.1 \\
{\LARGE {$\star$}} 
& $\zeta$     Tau &  HR 1910 & 2.88 &  0.29      & 24.7 \\
{\LARGE {$\star$}} 
& 78          Tau &  HR 1412 & 3.35 &  0.07      &  7.5 \\
{\normalsize $\ast$} 
& $\lambda$   Tau &  HR 1239 & 3.37 &  0.54      &      \\
{\large $\star$} 
& $\epsilon$  Tau &  HR 1409 & 3.53 &  $-$       &  9.4 \\
{\large $\star$} 
& $o$         Tau &  HR 1030 & 3.60 &  $-$       &  9.4 \\
{\large $\star$} 
& 27          Tau &  HR 1178 & 3.63 &  $-$       & 11.9 \\
{\large $\star$} 
& $\gamma$    Tau &  HR 1346 & 3.65 &  $-$       &  5.5 \\
{\large $\star$} 
& 17          Tau &  HR 1142 & 3.70 &  $-$       & 12.2 \\
{\large $\star$} 
& $\xi$       Tau &  HR 1038 & 3.70 &  0.09      &  8.7 \\
{\large $\star$} 
& $\delta$    Tau &  HR 1373 & 3.76 &  $-$       &  7.3 \\
{\LARGE $\bullet$}
& 1           Ori &  HR 1543 & 3.19 &  $-$       & 12.9 \\
{\normalsize {$\bullet$}}
& 3           Ori &  HR 1552 & 3.69 &  $-$       & 14.2 \\
\tableline
{\LARGE {$\star$}} 
& $\beta$     Gem &  HR 2990 & 1.14 &  $-$       & 12.4 \\
{\LARGE {$\star$}} 
& $\alpha$    Gem &  HR 2891 & 1.59 &  $-$       & 13.4 \\
{\LARGE {$\star$}} 
& $\gamma$    Gem &  HR 2421 & 1.93 &  $-$       &  7.5 \\
{\LARGE {$\star$}} 
& $\mu$       Gem &  HR 2286 & 2.75 &  0.27      &  9.8 \\
{\LARGE {$\star$}} 
& $\epsilon$  Gem &  HR 2473 & 2.98 &  $-$       &  6.5 \\
{\LARGE {$\star$}} 
& $\eta$      Gem &  HR 2216 & 3.15 &  0.75      & 11.6 \\
{\LARGE {$\star$}} 
& $\xi$       Gem &  HR 2484 & 3.36 &  $-$       &  8.7 \\
{\LARGE {$\star$}} 
& $\delta$    Gem &  HR 2777 & 3.53 &  $-$       &  4.0 \\
{\LARGE {$\star$}} 
& $\kappa$    Gem &  HR 2985 & 3.57 &  $-$       & 10.1 \\
{\LARGE {$\star$}} 
& $\lambda$   Gem &  HR 2763 & 3.58 &  $-$       &  5.2 \\
{\LARGE {$\star$}} 
& $\theta$    Gem &  HR 2540 & 3.60 &  $-$       & 13.7 \\
{\normalsize $\ast$} 
& $\zeta$     Gem &  HR 2650 & 3.62 &  0.56      &      \\
{\large $\star$} 
& $\iota$     Gem &  HR 2821 & 3.79 &  $-$       &  8.9 \\
{\large $\star$} 
& $\upsilon$  Gem &  HR 2905 & 4.06 &  $-$       &  9.7 \\
{\LARGE $\bullet$}
& $\beta$     CMi &  HR 2845 & 2.84 &  0.08      & 13.5 \\
\tableline
{\LARGE {$\star$}} 
& $\alpha$    Car &  HR 2326 &-0.72 &  $-$       & 27.8 \\
{\LARGE {$\star$}} 
& $\beta$     Car &  HR 3685 & 1.68 &  $-$       &  7.9 \\
{\LARGE {$\star$}} 
& $\epsilon$  Car &  HR 3307 & 1.86 &  $-$       & 10.4 \\
{\LARGE {$\star$}} 
& $\iota$     Car &  HR 3699 & 2.25 &  $-$       &  4.7 \\
{\LARGE {$\star$}} 
& $\theta$    Car &  HR 4199 & 2.76 &  $-$       &  6.7 \\
{\LARGE {$\star$}} 
& $\upsilon$  Car &  HR 3890 & 3.01 &  $-$       &  2.6 \\
{\LARGE {$\star$}} 
& $\omega$    Car &  HR 4037 & 3.32 &  $-$       &  8.0 \\
{\LARGE {$\star$}} 
& p           Car &  HR 4140 & 3.27 & 0.10       &  5.5 \\
{\large $\star$} 
& q           Car &  HR 4050 & 3.36 & 0.08       &  3.9 \\
{\large $\star$} 
& a           Car &  HR 3659 & 3.44 &  $-$       &  5.5 \\
{\large $\star$} 
& $\chi$      Car &  HR 3117 & 3.47 &  $-$       & 17.1 \\
{\normalsize $\ast$} 
& l           Car &  HR 3884 & 3.28 & 0.90       &      \\
{\large $\star$} 
& U           Car &  HR 4257 & 3.78 &  $-$       &  9.0 \\
{\large $\star$} 
& S           Car &  HR 4114 & 3.82 &  $-$       &  6.4 \\
{\large $\star$} 
& c           Car &  HR 3571 & 3.84 &  $-$       &  6.2 \\
{\large $\star$} 
& X           Car &  HR 4337 & 3.84 & 0.18       & 10.7 \\
{\large $\star$} 
& i           Car &  HR 3663 & 3.97 &  $-$       &  3.9 \\
{\large $\star$} 
& I           Car &  HR 4102 & 4.00 &  $-$       & 12.0 \\
{\LARGE $\bullet$}
& $\delta$    Vel &  HR 3485 & 1.96 & 0.40       & 11.1 \\
{\LARGE $\bullet$}
& $\kappa$    Vel &  HR 3734 & 2.50 &  $-$       &  8.1 \\
{\LARGE $\bullet$}
& n           Vel &  HR 3803 & 3.13 &  $-$       &  5.7 \\
{\normalsize {$\bullet$}}
& $\phi$      Vel &  HR 3940 & 3.54 &  $-$       &  8.1 \\
{\normalsize {$\bullet$}}
& $o$         Vel &  HR 3447 & 3.55 & 0.12       & 12.8 \\
{\LARGE $\bullet$}
& $\lambda$   Cen &  HR 4467 & 3.13 &  $-$       & 12.6 \\
{\normalsize {$\bullet$}}
& $\pi$       Cen &  HR 4390 & 3.89 & $-$        & 14.7 \\
{\normalsize {$\bullet$}}
& $\beta$     Vol &  HR 3347 & 3.77 &  $-$       &  9.3 \\
{\normalsize {$\bullet$}}
& $\alpha$    Vol &  HR 3615 & 4.00 &  $-$       &  6.0 \\
{\normalsize {$\bullet$}}
& $\lambda$   Mus &  HR 4520 & 3.64 &  $-$       & 13.4 \\
\tableline
\end{tabular}
\end{scriptsize}
\addtolength{\tabcolsep}{0.0cm}
\renewcommand{\arraystretch}{1.00}
\end{center}
\end{table}

\clearpage

\begin{table}
\begin{center}
\renewcommand{\arraystretch}{1.00}
\addtolength{\tabcolsep}{0.0cm}
\begin{scriptsize}
\begin{tabular}{cllccc}
\tableline\tableline
Notation &  Name           & Number  & $m$   & $\Delta m$ & $d~~$\\
         &                 &         & [mag] & [mag]      & [da] \\
\tableline
{\LARGE {$\star$}} 
& $\alpha$    Lyr &  HR 7001 &-0.02 & 0.09       &  5.9 \\
{\LARGE {$\star$}} 
& $\gamma$    Lyr &  HR 7178 & 3.24 &  $-$       &  2.0 \\
{\normalsize $\ast$} 
& $\beta$     Lyr &  HR 7106 & 3.25 & 1.11       &      \\
{\large $\star$} 
& R           Lyr &  HR 7157 & 3.88 & 1.12       & 10.6 \\
{\normalsize {$\bullet$}}
& $\mu$       Her &  HR 6623 & 3.42 &  $-$       & 14.3 \\
{\normalsize {$\bullet$}}
& $\xi$       Her &  HR 6703 & 3.70 &  $-$       & 12.0 \\
{\normalsize {$\bullet$}}
& $o$         Her &  HR 6779 & 3.80 & 0.07       & 10.2 \\
{\normalsize {$\bullet$}}
& 109         Her &  HR 6895 & 3.84 &  $-$       & 12.9 \\
{\normalsize {$\bullet$}}
& $\theta$    Her &  HR 6695 & 3.86 &  $-$       & 11.6 \\
{\normalsize {$\bullet$}}
& 110         Her &  HR 7061 & 4.19 &  $-$       & 12.6 \\
{\LARGE $\bullet$}
& $\beta$     Cyg &  HR 7417 & 3.08 &  $-$       & 10.3 \\
{\normalsize {$\bullet$}}
& $\eta$      Cyg &  HR 7615 & 3.39 &  $-$       & 13.8 \\
{\normalsize {$\bullet$}}
& $\chi$      Cyg &  HR 7564 & 3.30 & 10.9       & 12.7 \\
\tableline
{\LARGE {$\star$}} 
& $\alpha$    Aql &  HR 7557 & 0.77 &  $-$       &  7.6 \\
{\LARGE {$\star$}} 
& $\gamma$    Aql &  HR 7525 & 2.72 &  $-$       &  9.7 \\
{\LARGE {$\star$}} 
& $\zeta$     Aql &  HR 7235 & 2.99 &  $-$       & 17.4 \\
{\LARGE {$\star$}} 
& $\theta$    Aql &  HR 7710 & 3.23 &  $-$       &  4.7 \\
{\LARGE {$\star$}} 
& $\delta$    Aql &  HR 7377 & 3.36 &  $-$       &  7.2 \\
{\LARGE {$\star$}} 
& $\lambda$   Aql &  HR 7236 & 3.44 &  $-$       & 12.9 \\
{\large $\star$} 
& $\beta$     Aql &  HR 7602 & 3.71 &  $-$       &  5.9 \\
{\normalsize $\ast$} 
& $\eta$      Aql &  HR 7570 & 3.48 & 0.91       &      \\
{\large $\star$} 
& $\epsilon$  Aql &  HR 7176 & 4.02 &  $-$       & 19.2 \\
{\large $\star$} 
& i           Aql &  HR 7193 & 4.02 &  $-$       & 14.3 \\
{\LARGE $\bullet$}
& $\alpha$    Cap &  HR 7754 & 0.08 &  $-$       & 15.0 \\
{\normalsize {$\bullet$}}
& $\epsilon$  Del &  HR 7852 & 4.03 &  $-$       & 14.4 \\
\tableline
{\LARGE {$\star$}} 
& $\alpha$    Cep &  HR 8162 & 2.44 &  $-$       &  9.6 \\
{\LARGE {$\star$}} 
& $\gamma$    Cep &  HR 8974 & 3.21 &  $-$       & 20.0 \\
{\LARGE {$\star$}} 
& $\beta$     Cep &  HR 8238 & 3.16 & 0.11       & 13.7 \\
{\LARGE {$\star$}} 
& $\zeta$     Cep &  HR 8465 & 3.35 &  $-$       &  2.4 \\
{\LARGE {$\star$}} 
& $\eta$      Cep &  HR 7957 & 3.43 &  $-$       & 13.3 \\
{\LARGE {$\star$}} 
& $\iota$     Cep &  HR 8694 & 3.52 &  $-$       &  8.3 \\
{\normalsize $\ast$} 
& $\delta$    Cep &  HR 8571 & 3.48 & 0.89       &      \\
{\large $\star$} 
& $\mu$       Cep &  HR 8316 & 3.43 & 1.67       &  5.9 \\
{\large $\star$} 
& $\epsilon$  Cep &  HR 8494 & 4.15 & 0.06       &  2.5 \\
{\large $\star$} 
& $\theta$    Cep &  HR 7850 & 4.22 &  $-$       & 15.2 \\
{\LARGE $\bullet$}
& $\beta$     Cas &  HR 21   & 2.25 & 0.06       & 12.7 \\
{\normalsize {$\bullet$}}
& $\alpha$    Lac &  HR 8585 & 3.77 &  $-$       &  8.2 \\
{\normalsize {$\bullet$}}
& 81          Cyg &  HR 8335 & 4.23 &  $-$       & 11.0 \\
\tableline
\end{tabular}
\end{scriptsize}
\addtolength{\tabcolsep}{0.0cm}
\renewcommand{\arraystretch}{1.00}
\end{center}
\end{table}

\end{document}